\documentclass[pdflatex,sn-mathphys-num]{sn-jnl}

\usepackage{graphicx}%
\usepackage{multirow}%
\usepackage{amsmath,amssymb,amsfonts}%
\usepackage{amsthm}%
\usepackage{mathrsfs}%
\usepackage[title]{appendix}%
\usepackage{xcolor}%
\usepackage{textcomp}%
\usepackage{manyfoot}%
\usepackage{booktabs}%
\usepackage{algorithm}%
\usepackage{algorithmicx}%
\usepackage{algpseudocode}%
\usepackage{listings}%
\usepackage{xfrac}

\theoremstyle{thmstyleone}%
\theoremstyle{thmstyletwo}%

\theoremstyle{thmstylethree}%

\newcommand{\gev}{\ensuremath{\mathrm{GeV}}}
\newcommand{\taupi}{\ensuremath{\tau^\pm \rightarrow \pi^\pm \nu}}
\newcommand{\taurho}{\ensuremath{\tau^\pm \rightarrow \pi^\pm \pi^0 \nu}}

\begin{document}

\title[Article Title]{Experimental aspects of the Quantum Tomography of tau lepton pairs at a Higgs factory collider}


\author{\fnm{Daniel} \sur{Jeans}}\email{daniel.jeans@kek.jp}

\affil{\orgdiv{IPNS}, \orgname{KEK}, \orgaddress{\street{1-1 Oho}, \city{Tsukuba}, \country{Japan}}}


\abstract{
  Quantum Tomography of tau lepton pairs produced at a Higgs Factory collider
  will enable measurements of their spin correlations arising from quantum entanglement.
  Such measurements rely on the ability to measure the components of and correlations between the taus' spins.
  We present a method to fully reconstruct the kinematics of tau pair events at an electron-positron Higgs factory collider,
  making use of measured particles' momenta and impact parameters.
  This procedure results in several consistent solutions per event, which can be assigned weights according to various event properties.
  Full kinematic reconstruction allows the optimal extraction of the taus' spin orientation via polarimeters.
  We estimate the precision with which polarimeters can be reconstructed in an ideal detector,
  and quantify the effects of more realistic detector performance.
  We conclude that for this analysis, achieving a photon angular resolution of around 0.1~mrad is the
  most crucial aspect of detector performance, while photon energy resolution
  and vertex detector performance are significantly less important.
}


\maketitle

\section{Introduction}

The process $e^+ e^- \to \tau^+ \tau^- (\gamma)$ is a fundamental reaction in electron-positron colliders.
A future electron-positron Higgs Factory collider, such as the proposed
International Linear Collider (ILC)~\cite{Behnke:2013xla},
Linear Collider Facility~\cite{LinearCollider:2025lya},
Compact Linear Collider~\cite{Adli:2025swq},
Future Circular Collider~\cite{FCC:2025lpp},
and Circular Electron Positron Collider~\cite{CEPCStudyGroup:2023quu} projects,
will produce such events at higher energies and with larger statistics than before.

The prompt decay of taus brings sensitivity to their spin orientation via the distribution
of their decay products. This allows measurements of e.g. the longitudinal tau polarisation
see e.g.~\cite{DELPHI:1999yne,DELPHI:2007ywu,Yumino:2022vqt},
which in turn provides
sensitivity to the fundamental Electro-Weak parameter $\sin \theta_W$.

Recently, there is much interest in the measurement of entangled quantum states at
colliders~\cite{Fabbrichesi:2022ovb,Ehataht:2023zzt,Fabbrichesi:2024wcd,Altakach:2026fpl,Guo:2026yhz,Ma:2023yvd,LoChiatto:2024dmx,Han:2025ewp}.
$e^+ e^- \to \tau^+ \tau^-$ is a suitable system in which to make such measurements, thanks to
the ability to reconstruct the spins of the two co-produced taus.

In this paper, we discuss various experimental aspects of the Quantum Tomography, {\em i.e.} the reconstruction of tau spin information, 
of the process $e^+ e^- \to \tau^+ \tau^- (\gamma)$.
Section~\ref{sec:reco} describes a method to reconstruct kinematic information from the
available measurements.
We present results assuming a detector with perfect performance at a 250~\gev\ ILC in Sec.~\ref{sec:ILCres}, and
the extraction of spin information is discussed in Sec.~\ref{sec:spin}.
Section~\ref{sec:detres} describes the use of fast simulation of the International Large Detector (ILD)~\cite{ILDConceptGroup:2020sfq}
to investigate the effects of varying the resolution of key detector components.


\section{Kinematic reconstruction}
\label{sec:reco}

The kinematic reconstruction of $e^+ e^- \to \tau^+ \tau^- (\gamma)$ presents several challenges.
The final state contains several undetectable particles: the neutrinos produced in the $\tau$ decay
-- two per $\tau$ for leptonic and one for semi-leptonic decays, 
and often initial state radiation (ISR) collinear to the beams.
Optimal reconstruction of spin information in $\tau$ decays requires knowledge of
all its daughter particle momenta, including the neutrinos.

Compared to our previous analysis of tau reconstruction in 
$e^+ e^- \to \mu^+ \mu^- \tau^+ \tau^- (\gamma)$ ({\em e.g.} via 
$e^+ e^- \to Z H$)~\cite{Jeans:2015vaa,Jeans:2018anq},
the current process is less constrained due the absence of an independent measurement of the
interaction point (IP) provided by the two muons. Whereas the previous approach
considered only the transverse momentum balance in an event, thereby rendering it insensitive to
the emission of collinear ISR, in this case we must explicitly consider ISR.

\begin{figure}
\begin{center}
\includegraphics[width=0.4\textwidth]{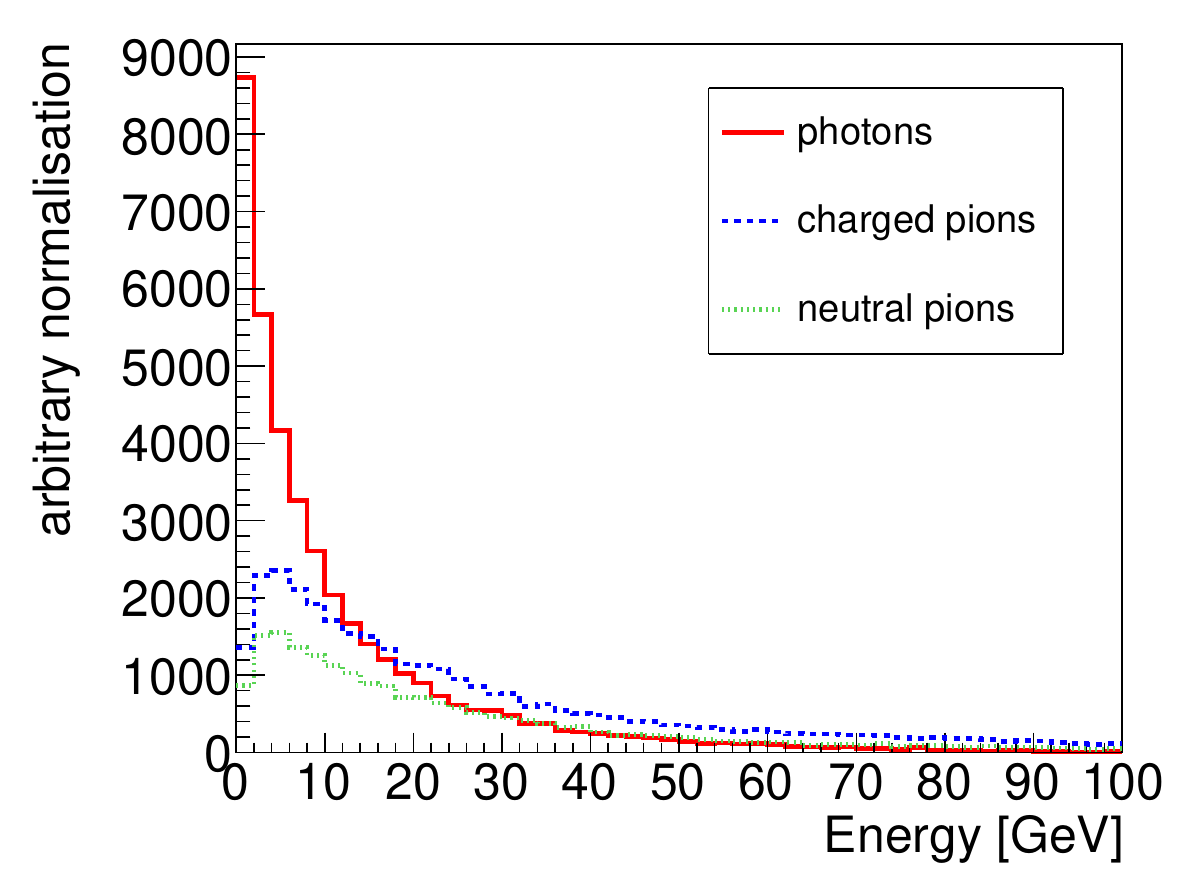}
\caption{
  Distribution of pion and photon energies in the considered decay modes (right) at 250~\gev\ ILC.
}
\label{fig:piens}
\end{center}
\end{figure}

We restrict ourselves to events in which both taus decay semi-leptonically.
This subsample of events is a more tractable system, containing only two neutrinos per event.
The spin sensitivity in semi-leptonic decays is significantly larger than that of leptonic decays,
due to the presence of two neutrinos in the latter.
The sensitivity is in principle uniform amongst semi-leptonic decay channels,
provided the form of the hadronic current is known.
The hadronic current, a function of the momenta of tau decay daughters,
can be used to extract the polarimeter, an optimal estimator of the spin orientation.
We consider tau decays to $\pi^\pm\nu$, $\pi^\pm\pi^0\nu$, $\pi^\pm\pi^0\pi^0\nu$, and $\pi^\pm\pi^\pm\pi^\mp\nu$,
respectively accounting for 10.8\%, 25.5\%, 9.3\% and 9.3\% of tau decays,
for which optimal polarimeters can relatively easily be calculated.
The energy of the tau decay products in these decays is shown in Fig.~\ref{fig:piens}.

The visible products of these tau decay modes ideally consist of only charged pions and photons
(from either neutral pion decay or visible ISR).
The photons are reconstructed as clusters within the electromagnetic calorimeter, allowing an estimate of their
energy and direction.
The momentum of charged pions is measured in the tracking detector, typically placed within a solenoidal magnetic field.
Since the tau has a mean flight distance of a few mm
($\beta \gamma c \tau \sim 5\textrm{mm}$ for a 100~\gev\ tau),
the charged pion trajectories are typically significantly displaced from the interaction point (IP).
The experiment's vertex detector can measure the pion's trajectory with a precision of a few microns in the vicinity of the IP.
Since all charged particles produced in the event are displaced, no direct measurement of the IP position is available,
beyond knowledge of the envelope of the collider's luminous region.

\begin{figure}
\begin{center}
\includegraphics[width=0.4\textwidth]{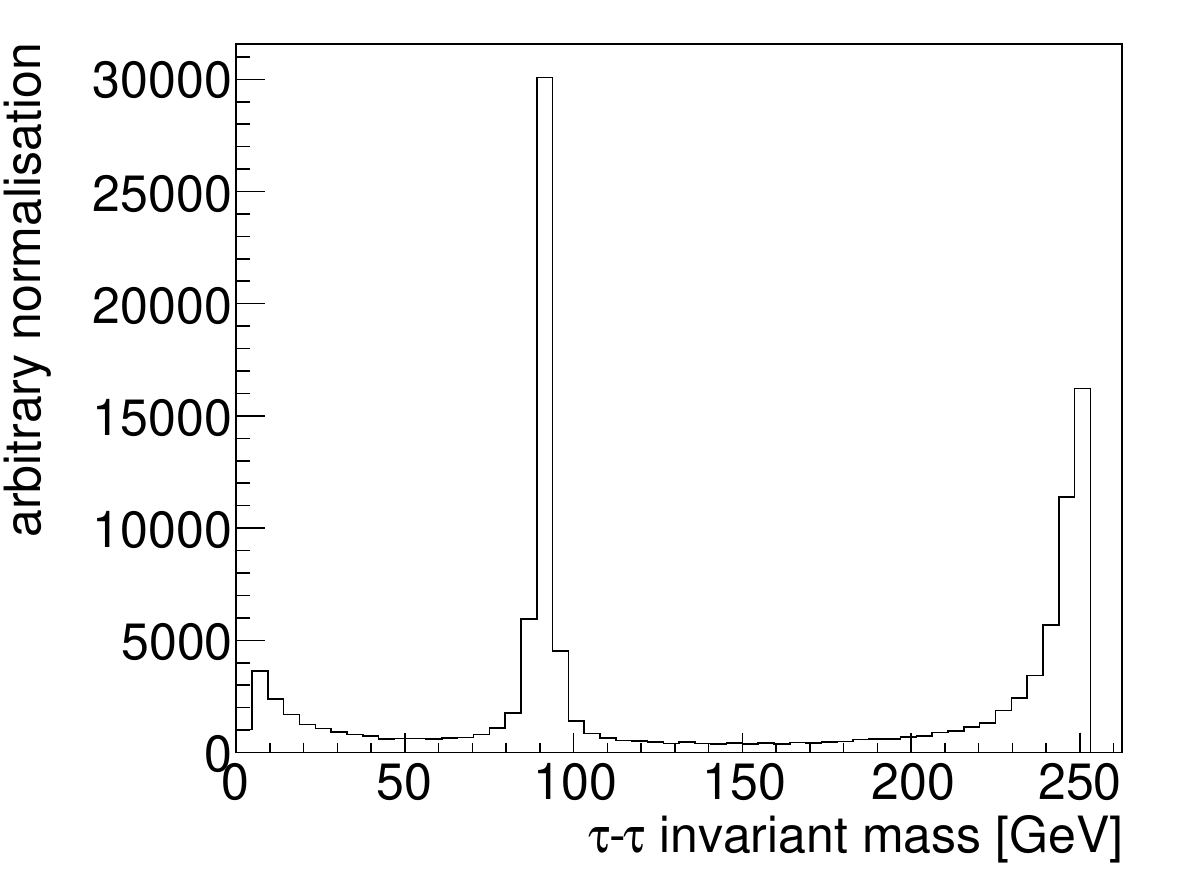}
\caption{
  Simulated distribution of the $\tau^- \tau^+$ invariant mass at 250~\gev\ ILC.
}
\label{fig:mtt}
\end{center}
\end{figure}

The centre-of-mass (CoM) energy is quite precisely known at an electron-positron collider, however
a complication that cannot be ignored is ISR, in which one or both of the initial $e^+ e^-$
radiates a photon before the hard collision.
This photon is preferentially emitted in the
beam direction, and therefore often escapes detection, passing through the outgoing beampipe.
The effect of ISR on the invariant mass of the tau pair is demonstrated in Fig.~\ref{fig:mtt},
which shows that this mass is significantly less than the nominal CoM energy for the majority of events.
The radiative return to the Z boson mass is clearly seen.

The tau mass $m_\tau$, while not directly observable due to its decay into at least one neutrino, can be used as a constraint
on the event kinematics. Since $m_\tau$ is much smaller than the typical tau energy $\mathcal{O}(100)~\gev$,
we may expect that such constraints are rather sensitive to even small mis-measurements of 
particle energies and directions as well as to variations in the initial conditions due to e.g. beam energy spread.

In the following we outline a method to reconstruct the full kinematics of $e^+ e^- \to \tau^+ \tau^- (\gamma)$ events,
taking the above considerations into account.

\subsubsection*{Non-tau objects}
In general, there may be visible activity from the hard event which is not associated with the tau-tau system,
for example visible ISR photons.
We define its 4-momentum in the laboratory (LAB) frame as $p_{OTHER}$.

It is likely that ISR may have been emitted from one or both initial beam particles,
and evaded detection due to the preference for emission collinear with the initial beam direction.
In this analysis we will assume that either one or both beams have emitted collinear photons,
with total 4-momentum $p_{ISR}$ of the ISR system.
We define these in the LAB frame, taking account of any collider crossing angle.
Since we cannot know {\em a priori} the energy of these photons, we will scan across different possibilities,
and test which are consistent with the measured event properties and assumed event kinematics.

\subsubsection*{Tau--tau centre-of-mass frame}

Assuming the nominal beam momenta $P^\pm_{BEAM}$ and the resulting CoM 4-momentum
$P_{CoM} = P^+_{BEAM} + P^-_{BEAM}$ in the LAB frame,  
we write the 4-momentum of the system which recoils against the (OTHER + ISR) system: $P_{REC} = P_{CoM} - p_{OTHER} - p_{ISR}$.
This recoiling frame corresponds to the rest frame of the tau-tau system, if the ISR assumptions we have made are correct.
$M = P_{REC}^2$ is the invariant mass of this recoiling system.

We next boost the visible tau jet momenta from the LAB frame ($p_{VIS}$)
to the recoiling frame, assumed to be the tau-tau system's rest frame (we use $'$ to denote quantities in this recoiling frame: $p'_{VIS}$).
In this frame, the two tau momenta are by definition back-to-back, and each tau has energy $E'_\tau=M/2$.

\subsubsection*{Tau momentum: cone method}

Since we consider only semi-leptonic tau decays, 
a single neutrino of 4--momentum $[q', \mathbf{q'}]$ is produced in each tau decay.
We write this 4--momentum in terms of the tau and visible tau jet momenta:
\begin{equation}
  [q', \mathbf{q'}] = [E'_\tau, \mathbf{p'}_\tau ] - [E'_{VIS}, \mathbf{p}'_{VIS}].
\end{equation}
Using the massless neutrino property $m_\nu = 0$ and imposing the tau mass $m_\tau$,
\begin{eqnarray}
  0 = & m_\nu^2 \nonumber \\
   = & (E'_\tau - E'_{VIS})^2 - ( \mathbf{p'}_\tau - \mathbf{p'}_{VIS} )^2 \nonumber \\
  = & m_\tau^2 + m_{VIS}^2 - 2 E'_\tau E'_{VIS} + 2 | \mathbf{p'}_\tau | |\mathbf{p'}_{VIS} | \cos{\alpha} 
\end{eqnarray}
where $m_{VIS}$ is the invariant mass of the tau's visible decay products, and $\alpha$ the angle between the
the visible momentum and the unknown tau momentum.
The tau momentum $\mathbf{p'}_\tau$ must lie on a cone with half-angle $\alpha$ around the visible momentum direction $\mathbf{p'}_{VIS}$.
Setting the tau energy $E'_\tau = M/2$, we can write
\begin{align}
\cos\alpha & = \frac{ E'_{VIS} M - m_\tau^2 - m_{VIS}^2 }{ \sqrt{ ({E'_{VIS}}^2 - m_{VIS}^2 ) (M^2 - 4 m_\tau^2) } }.
\label{eqn:cosa}
\end{align}
Provided this expression is real and has a magnitude no larger than unity,
there exists a set of consistent solutions for the tau momentum.

\subsubsection*{Tau pair}

An event contains two taus, and therefore two cones defined in the tau-tau rest frame.
Since the taus are by definition back-to-back in this frame,
consistent tau-tau directions correspond to the lines along which
one cone -- after flipping the sign of its axis -- intersects with the
other cone. In general, the two cones intersect along zero or two directions.
Two cones with a common vertex at the origin,
normalised axis directions
($\mathbf{a} = \mathbf{\hat{p'}}^{+}_{VIS}, \mathbf{b} = -\mathbf{\hat{p'}}^{-}_{VIS}$)
separated by an angle
$\theta = \cos^{-1}(\mathbf{a \cdot b})$,
and respective half-angles
($\alpha, \beta$) as calculated from Eq.~\ref{eqn:cosa},
intersect along directions $\mathbf{r}_\pm$ given by
\begin{equation}
\mathbf{r}_\pm = \mathbf{c} \times (\cos{\beta}\ \mathbf{a} - \cos{\alpha}\ \mathbf{b}) \pm \kappa \mathbf{c}
\end{equation}
where $ \mathbf{c} = \mathbf{a} \times \mathbf{b} $ and
\begin{equation}
\kappa^2 = 1 - \cos^2\alpha - \cos^2\beta - \cos^2\theta + 2 \cos\alpha \cos\beta \cos\theta\ . \label{eqn:k2}
\end{equation}
The cones intersect if $\kappa^2 \geq 0$, in which case the two directions $\mathbf{r}_\pm$ define
two possible solutions for the tau momentum direction in the tau-tau rest frame.
Since we know the energy $(E'_\tau = M/2)$ and mass of the taus in this frame, their 4-momenta are trivially derived.
The tau 4-momenta $p'_\tau$ are then boosted back to the laboratory frame ($p''_\tau$).

\subsubsection*{Scanning over $p_{ISR}$}

To arrive at a set of possible solutions, we scan over possible ISR scenarios.
We first assume a single ISR photon, collinear with one of the beams, and scan over all possible momenta.
In some regions of $p_{ISR}$, one or both taus may not have a real cone solution (Eq.~\ref{eqn:cosa});
even if both cone solutions are real, they may not intersect (Eq.~\ref{eqn:k2}).
If no intersecting cone solution is found, we allow both beams to emit a collinear photon,
scanning over possible energies for both beams' ISR.
The region of $p_{ISR}$ in which solutions are valid is typically quite limited; we scan across
100 evenly spaced points within the valid range of each beam's ISR, providing a set of possible solutions.
There is no general way to decide which of these solutions is correct, so we consider various properties
of the solutions to assign each a weight.

\subsubsection*{Weighting solutions: impact parameter information}

To decide whether a particular solution
-- defined by the choice of initial ISR momentum and one of the two cone intersections --
is valid, we can make use of the trajectories of charged tau decay products.

We first consider the case of tau decays to a single charged particle
and one or more neutral particles: the neutrino and possibly neutral pions.
Under the assumption that charged particle trajectory is linear at the scale being considered
(the decay length of the tau),
a general point $\mathbf{Q}_p$ on the charged particle trajectory can be written as
\begin{equation}
\mathbf{Q}_p = \mathbf{d}_{PCA} + \sigma \mathbf{p}_{PCA}
\label{eqn:chgTraj}
\end{equation}
where $\mathbf{d}_{PCA}$ is some point on the trajectory (e.g. the point of closest approach to the nominal IP),
$\mathbf{p}_{PCA}$ is the momentum at that point, and $\sigma \in \mathbb{R}$.

The transverse size of the luminous region at $e^+e^-$ Higgs factories is typically smaller than the
track impact parameter resolution, while its length along the beamline is much larger than
the experimental resolution.
The unknown production point of the taus is assumed to lie somewhere along the nominal beamline, at $x=y=0, z = z_{IP}$.
A general point $\mathbf{Q}_\tau$ along the unseen tau 
trajectory
(also assumed to be linear) can then be written as 
\begin{equation}
\mathbf{Q}_\tau = (0,0,z_{IP}) + \rho \mathbf{p}''_\tau
\label{eqn:tauTraj}
\end{equation}
where $\mathbf{p}''_\tau$ is the hypothesised tau momentum, and $\rho \in \mathbb{R}$.
We note that occasionally the tau flight distance will be sufficient to reach the vertex detector. This will
provide important additional information, however we do not consider it here.

For a consistent solution, the charged particle ($\mathbf{Q}_p$) and tau ($\mathbf{Q}_\tau$)
trajectories should meet at a point: the location of tau decay.
The two trajectories touch when
\begin{equation}
z_{IP} = ( \mathbf{f} \cdot \mathbf{d}_{PCA} ) / f_z 
\end{equation}
where
$\mathbf{f} = \mathbf{p}''_\tau \times \mathbf{p}_{PCA}$ and $f_z$ is its component along $z$.

For each candidate solution we thus get two values for $z_{IP}$, one per tau.
If they are consistent -- i.e. the two taus were produced at the same position along the beamline -- the solution is reasonable.

Tau decays with multiple charged decay products, such as the $\pi^\pm\pi^\pm\pi^\mp\nu$ decay we consider,
allow the explicit reconstruction of the tau decay vertex.
However, the error ellipsoid of the vertex's measured position is typically rather long
in one direction, due the small opening angle of the tau jet. We therefore treat
this case as a single trajectory which passes through the ellipsoid's center and is aligned with its major axis,
and follow the same procedure as above.

The two taus were produced at the same point, so their $z_{IP}$ can be compared to the track uncertainties on the
$z$ impact parameter $\delta z$ to estimate a likelihood that they are consistent:
\begin{equation}
\mathcal{L}_{dz} \propto \exp \bigg( - \frac{1}{2} \frac{ (z^+_{IP} - z^-_{IP})^2}{(\delta z^+)^2 + (\delta z^-)^2} \bigg).
\end{equation}
The average value of $z_{IP}$ can also be compared to the length of the luminous region along $z$, $\Delta z_{LUM}$
\begin{equation}
  \mathcal{L}_{z} \propto \exp \bigg( - \frac{1}{2} \frac{<z_{IP}>^2}{\Delta z_{LUM}^2 } \bigg),
\end{equation}
where $\Delta z_{LUM}$ is typically $\mathcal{O}(100 \mu m)$, depending on the collider being considered.

\subsubsection*{Weighting solutions: lifetime}

After this analysis of the impact parameters, we can calculate the reconstructed decay lengths of both taus.
If one or both are negative (i.e. $\rho$ of Eq.~\ref{eqn:tauTraj} is negative at the intersection) the solution is discarded.
If they are positive, we use the tau energy to calculate the taus' real lifetime $t^\pm$.
We define a likelihood that the calculated lifetimes are consistent with the tau mean lifetime $t_\tau = 0.29$~ps,
\begin{equation}
\mathcal{L}_{LIFE} \propto \exp( - t^+ / t_\tau  ) \cdot \exp( - t^- / t_\tau  ) .
\end{equation}

\subsubsection*{Weighting solutions: ISR}

The energy lost to ISR in $e^+ e^- \to \tau^+ \tau^- (\gamma)$ depends on the
spectrum of photon radiation from the beams and the CoM energy dependence of the $e^+ e^- \to \tau^+ \tau^-$ cross-section.
We create a probability density function (PDF) from the energy of ISR photons present in the generated signal event sample.
Each event solution has a particular assumed $p_{ISR}$ from either one or both beams,
based on which we assign a likelihood $\mathcal{L}_{ISR}$ from this PDF.

\subsubsection*{Weighting solutions: combination}

We assign to each solution a combined likelihood
$\mathcal{L} = \mathcal{L}_{dz} \cdot \mathcal{L}_{z} \cdot \mathcal{L}_{LIFE} \cdot \mathcal{L}_{ISR}$.
We reject solutions with likelihood less than 5\% of the event's highest likelihood solution.
Of the remaining solutions, we select a maximum of twenty solutions, evenly distributed along
the valid range of $p_{ISR}$. Solutions are assigned a weight proportional to their likelihood,
normalised so that each event has a total weight of unity. When histogramming an observable,
each of these solutions are included, with suitable weights.

\section{Application to tau-pair production at ILC-250.}
\label{sec:ILCres}

We apply this method to $e^+ e^- \to \tau^+ \tau^- (\gamma)$ events at the ILC at a CoM energy of 250~\gev.
Events were simulated using the WHIZARD event generator~\cite{Kilian:2007gr},
modeling the effects of both ISR and the expected colliding beam energy spectrum at ILC,
smeared due to both the beam energy spread and beamstrahlung.
Taus were decayed by TAUOLA~\cite{Jadach:1990mz, Jadach:1993hs}.
The analysed samples correspond to $40~\mathrm{pb^{-1}}$ of unpolarised collisions,
close to 0.4~M events in total.
We note that in these samples care was taken to accurately model
correlations between the longitudinal tau spin components, however
correlations between other components may less reliable.
This does not affect the results of this paper,
which concern the {\em precision} with which these components can be extracted.
The simulated distribution of the $\tau^- \tau^+$ invariant mass is shown in Fig.~\ref{fig:mtt}.
The population of events is enhanced around the nominal CM energy (250~\gev), the Z boson mass,
and around twice the tau mass (dominated by photon exchange).

\begin{figure*}[htb]
\begin{center}
\includegraphics[width=0.3\textwidth]{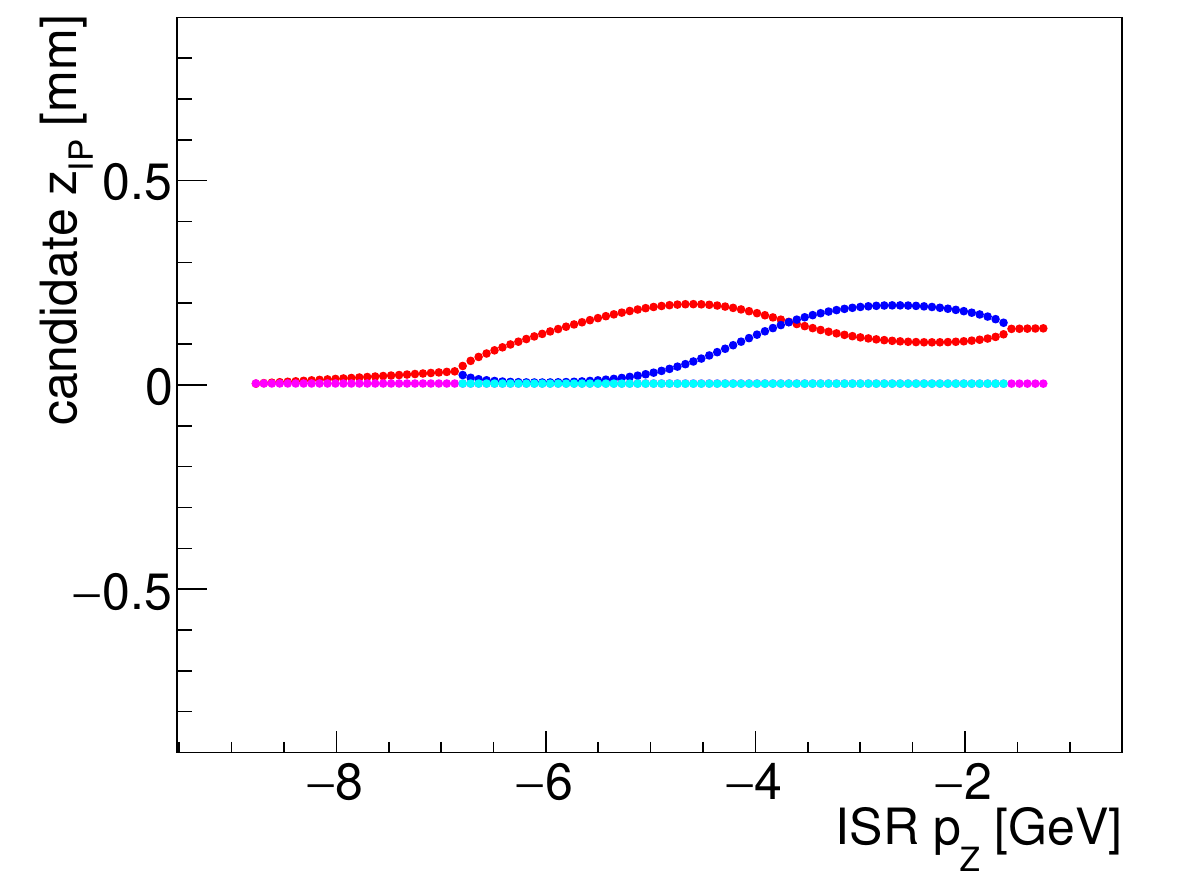}
\includegraphics[width=0.3\textwidth]{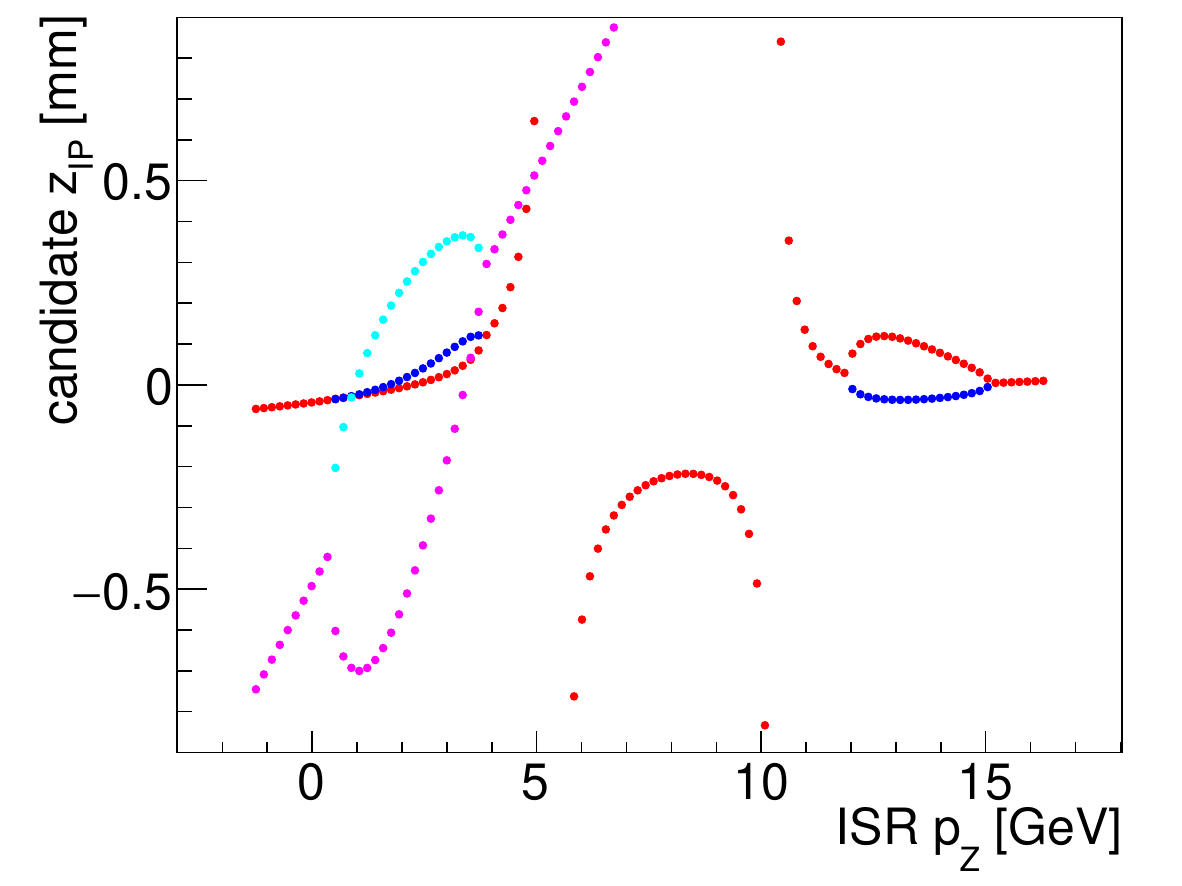}
\includegraphics[width=0.3\textwidth]{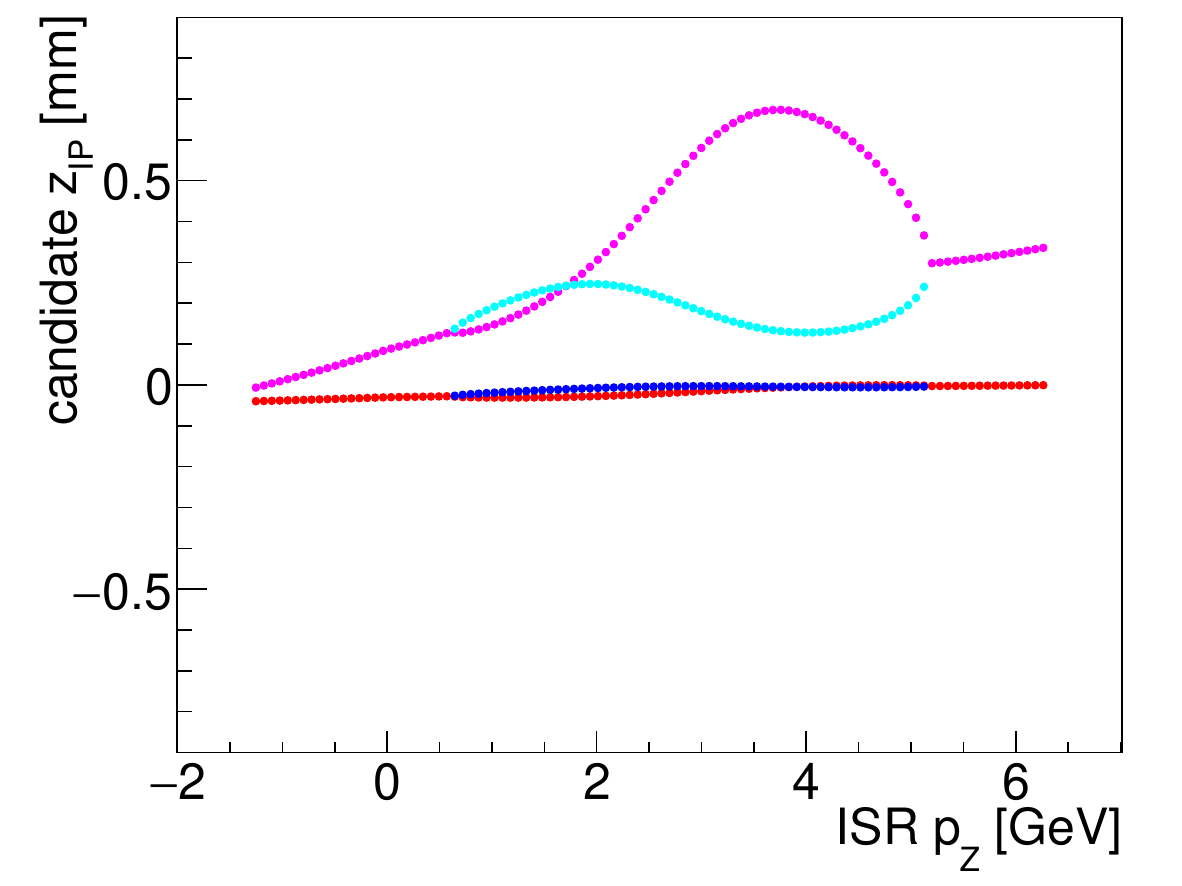} \\
\includegraphics[width=0.3\textwidth]{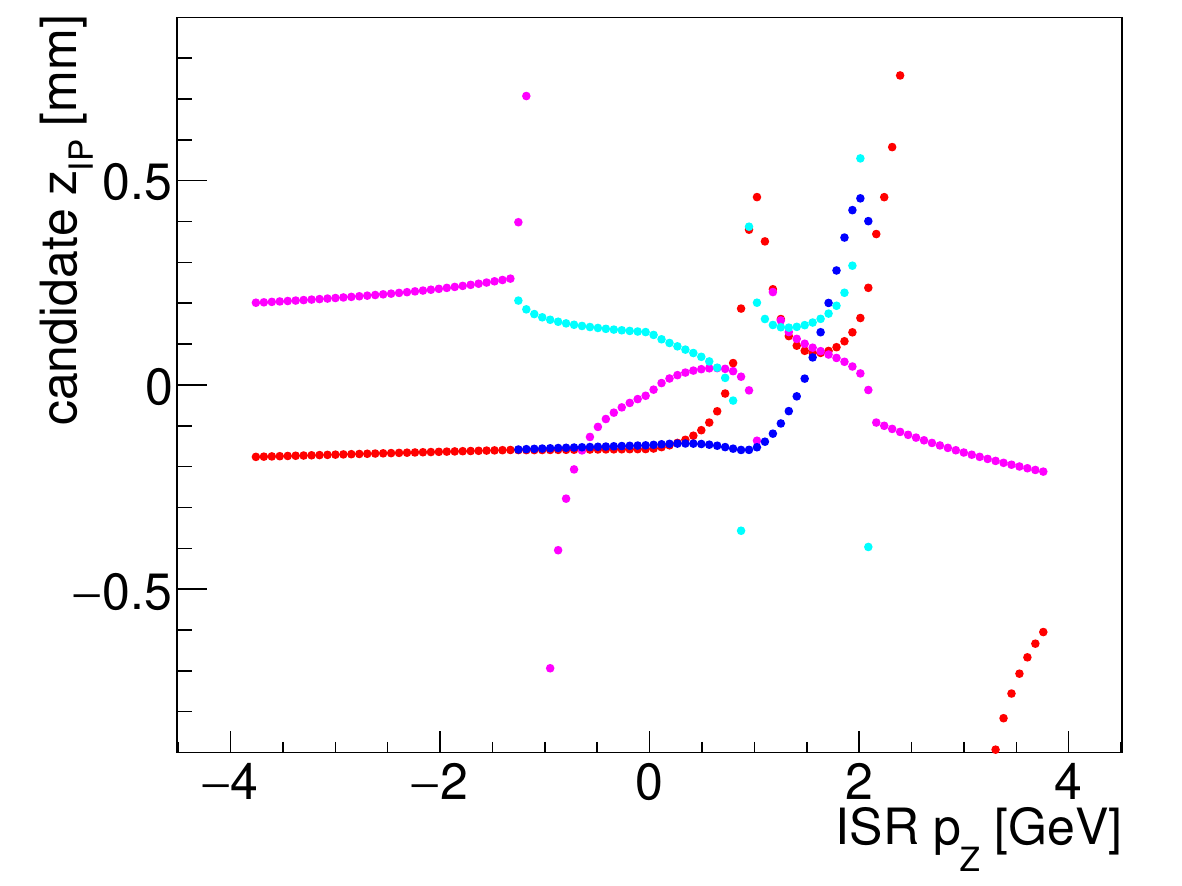} 
\includegraphics[width=0.3\textwidth]{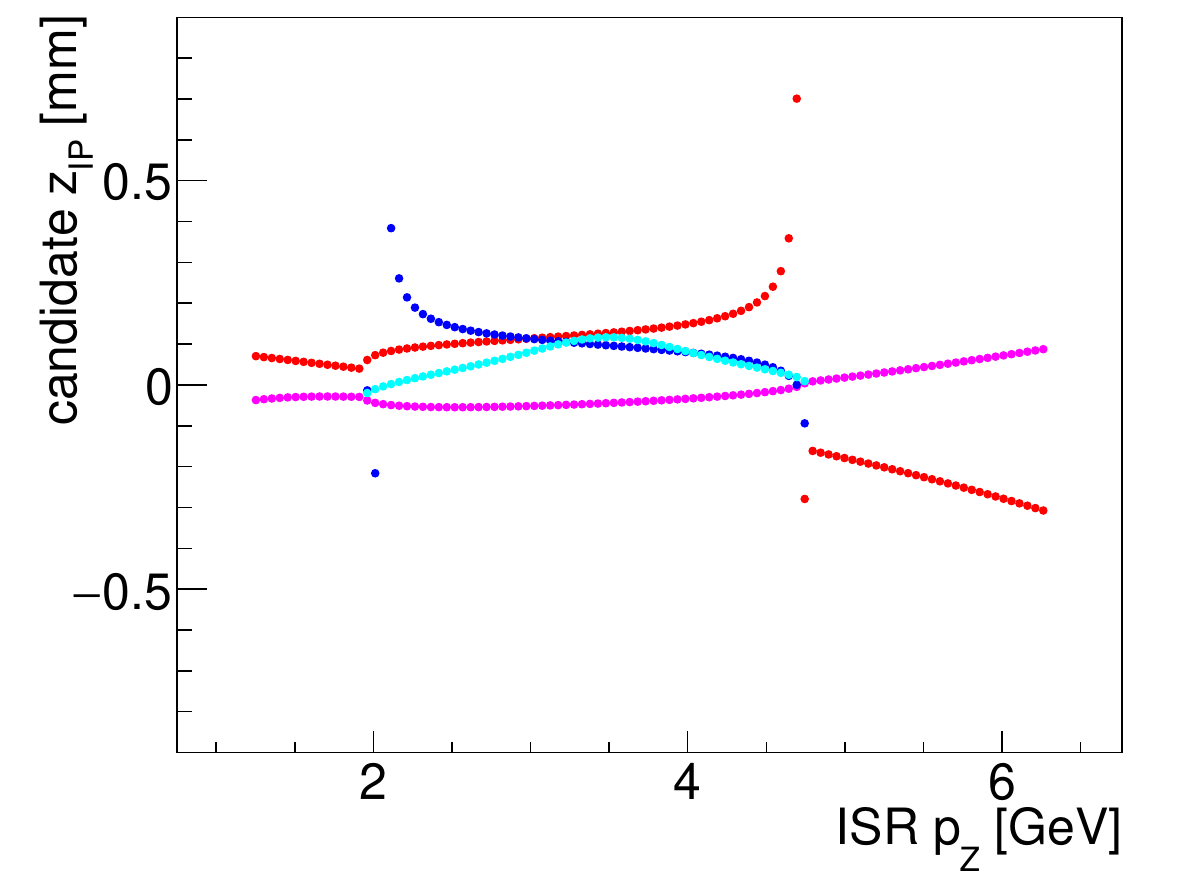}
\includegraphics[width=0.3\textwidth]{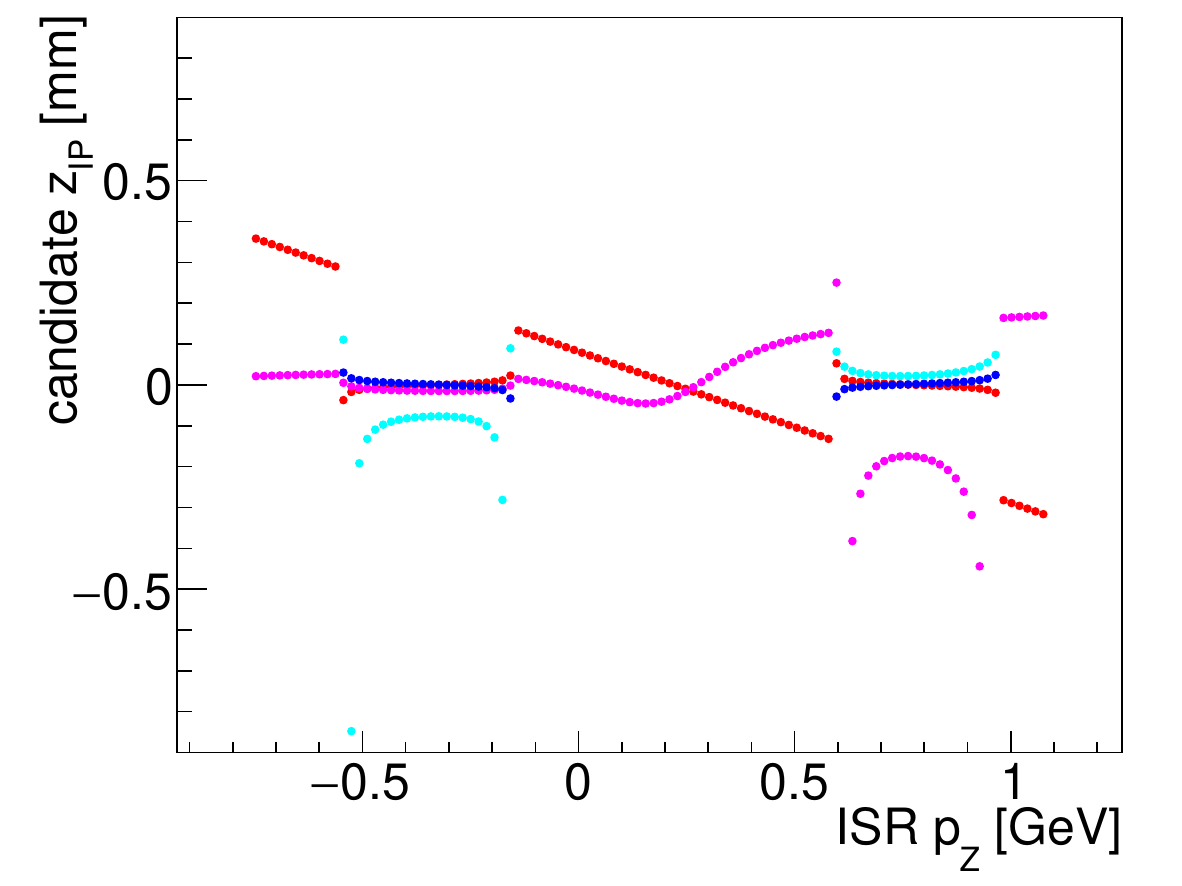} \\
\vspace{1cm}
\includegraphics[width=0.3\textwidth]{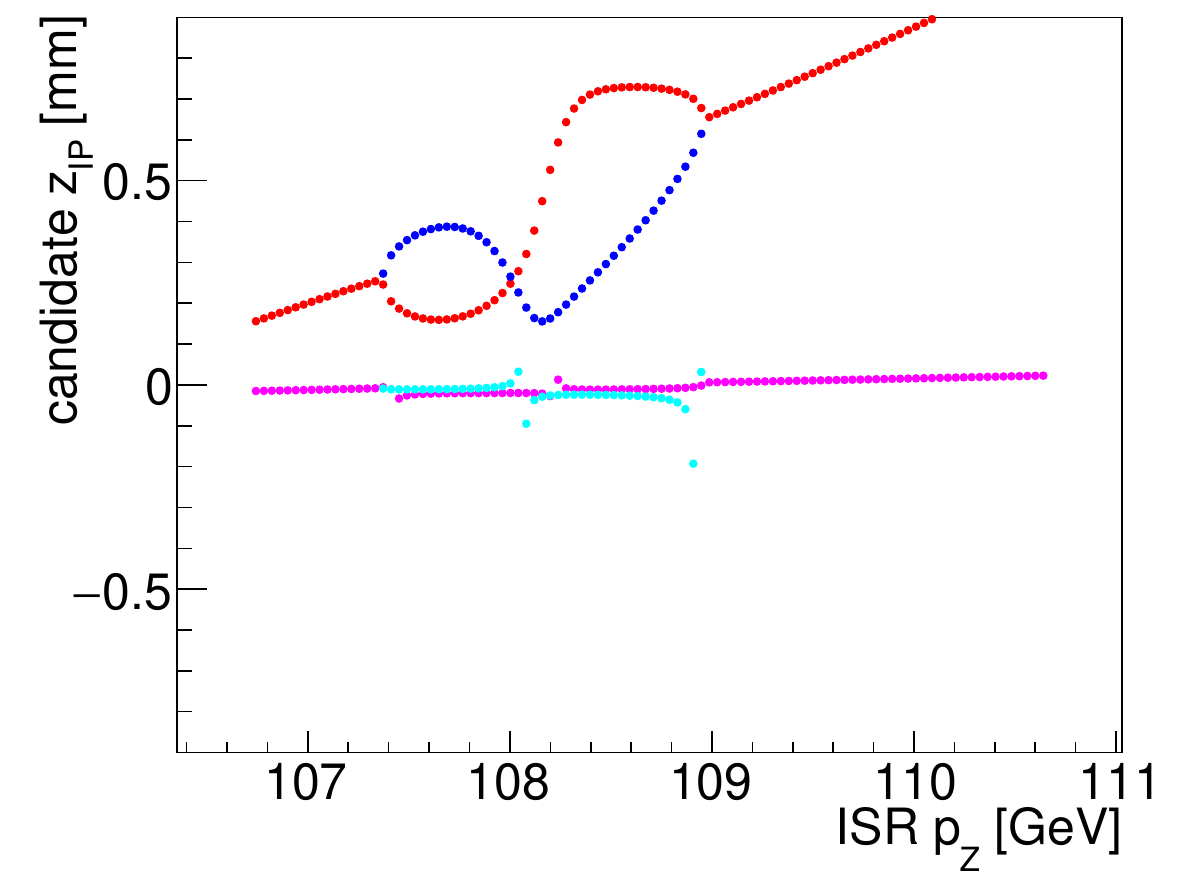}
\includegraphics[width=0.3\textwidth]{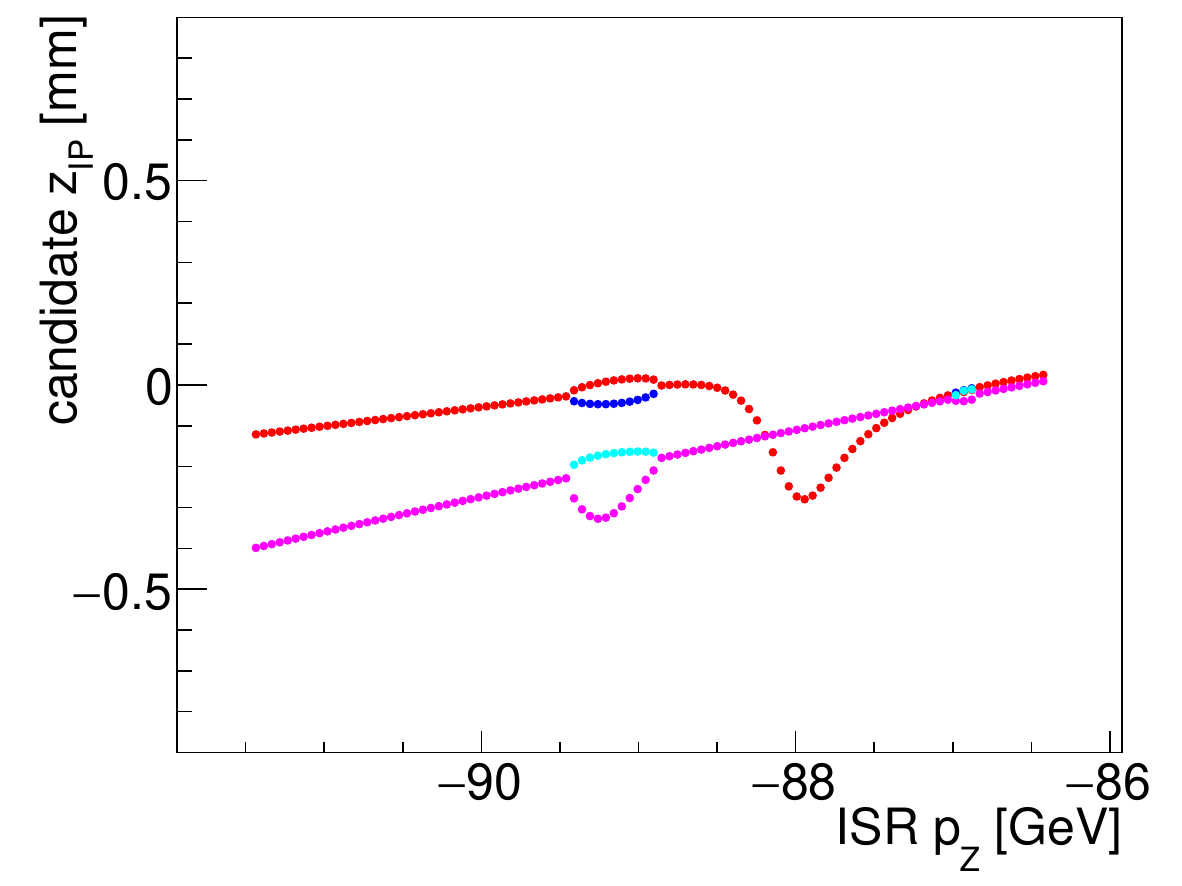}
\includegraphics[width=0.3\textwidth]{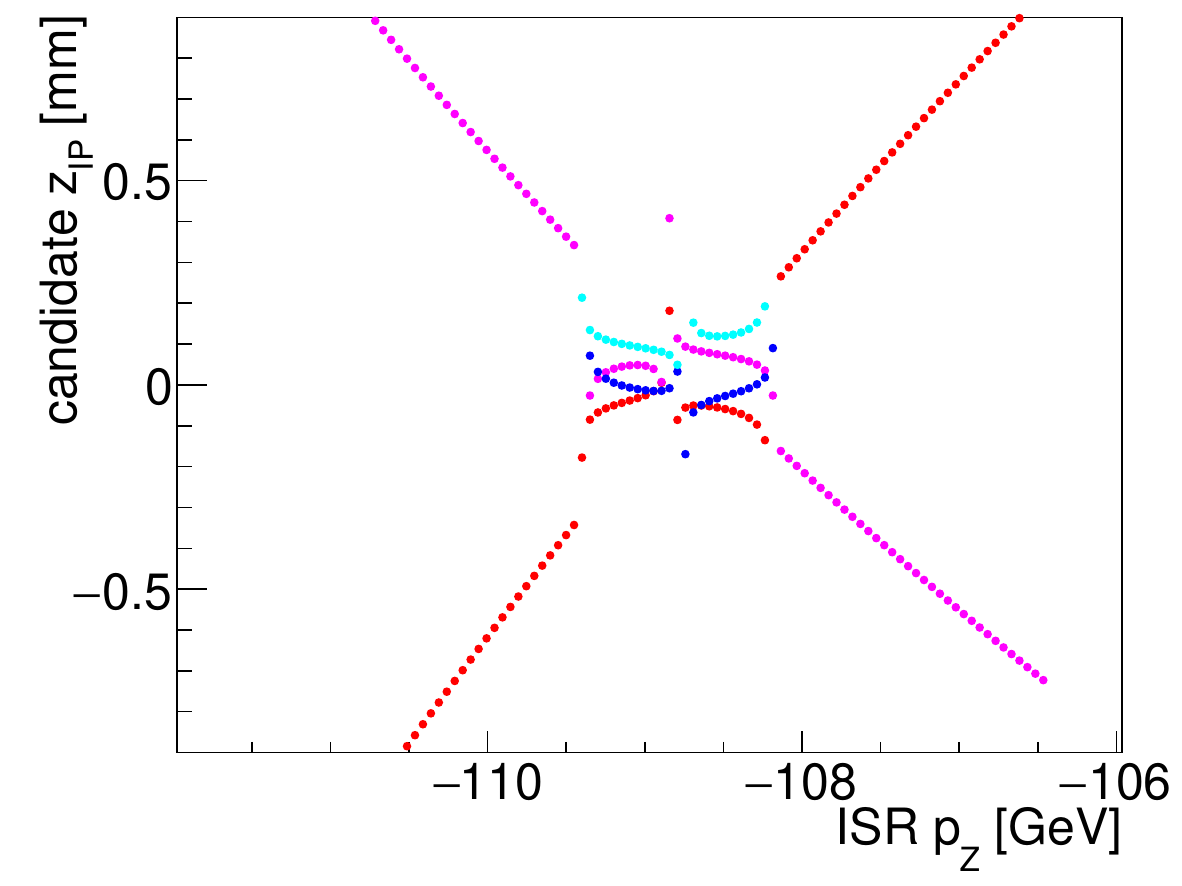} \\
\includegraphics[width=0.3\textwidth]{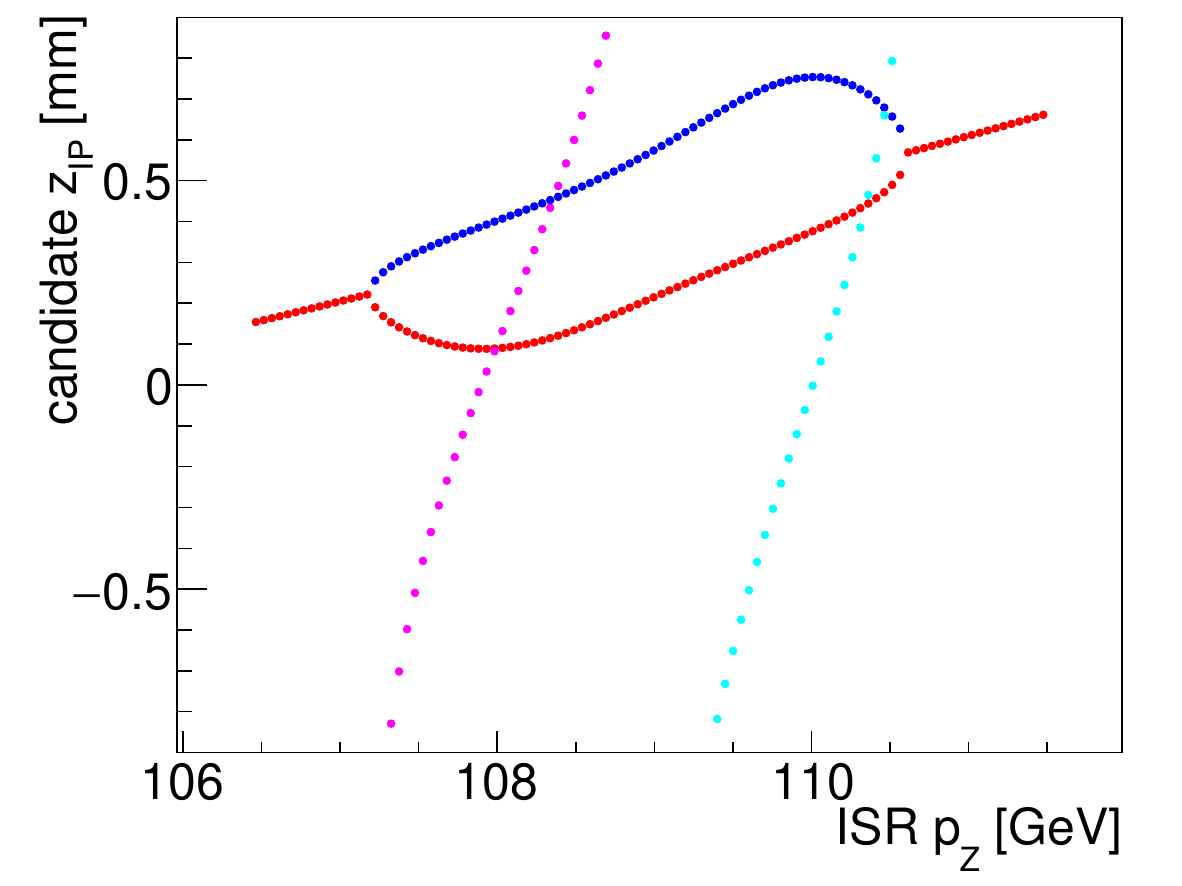}
\includegraphics[width=0.3\textwidth]{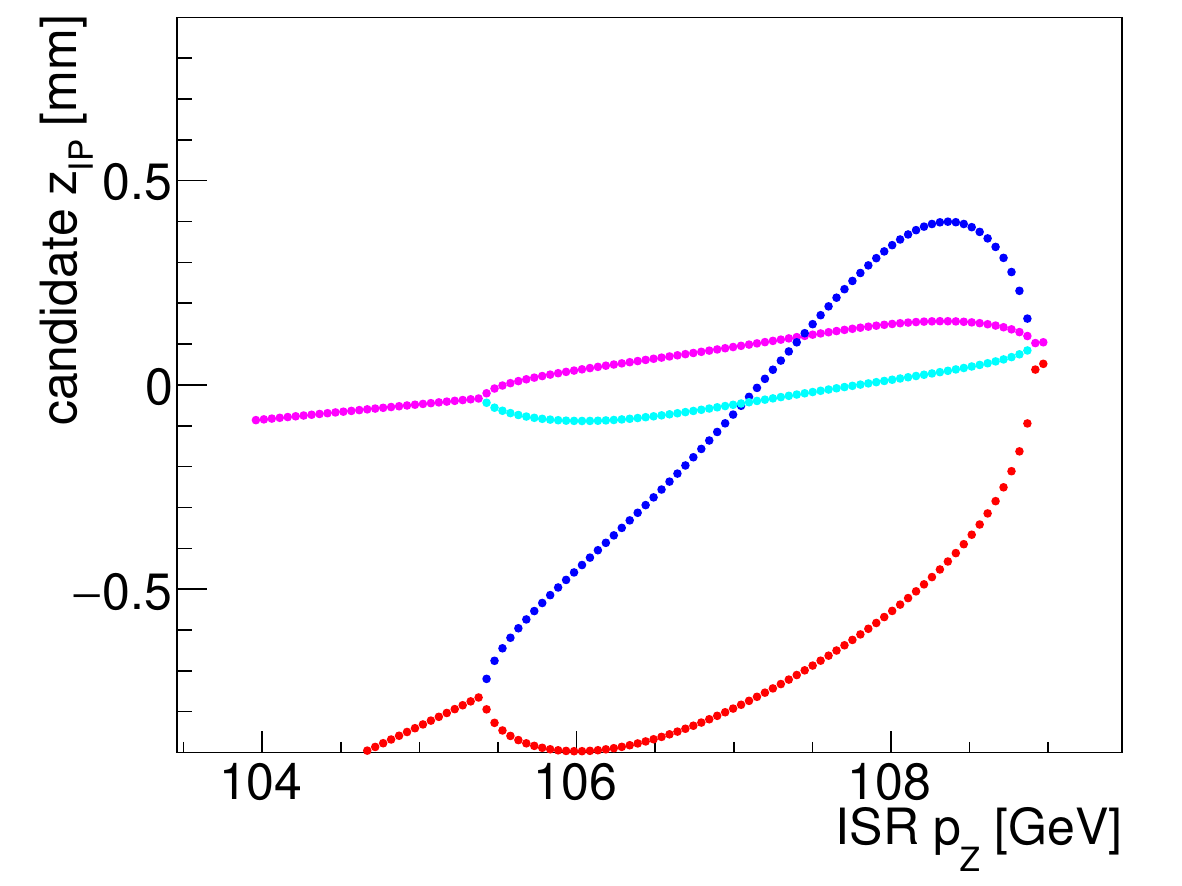}
\includegraphics[width=0.3\textwidth]{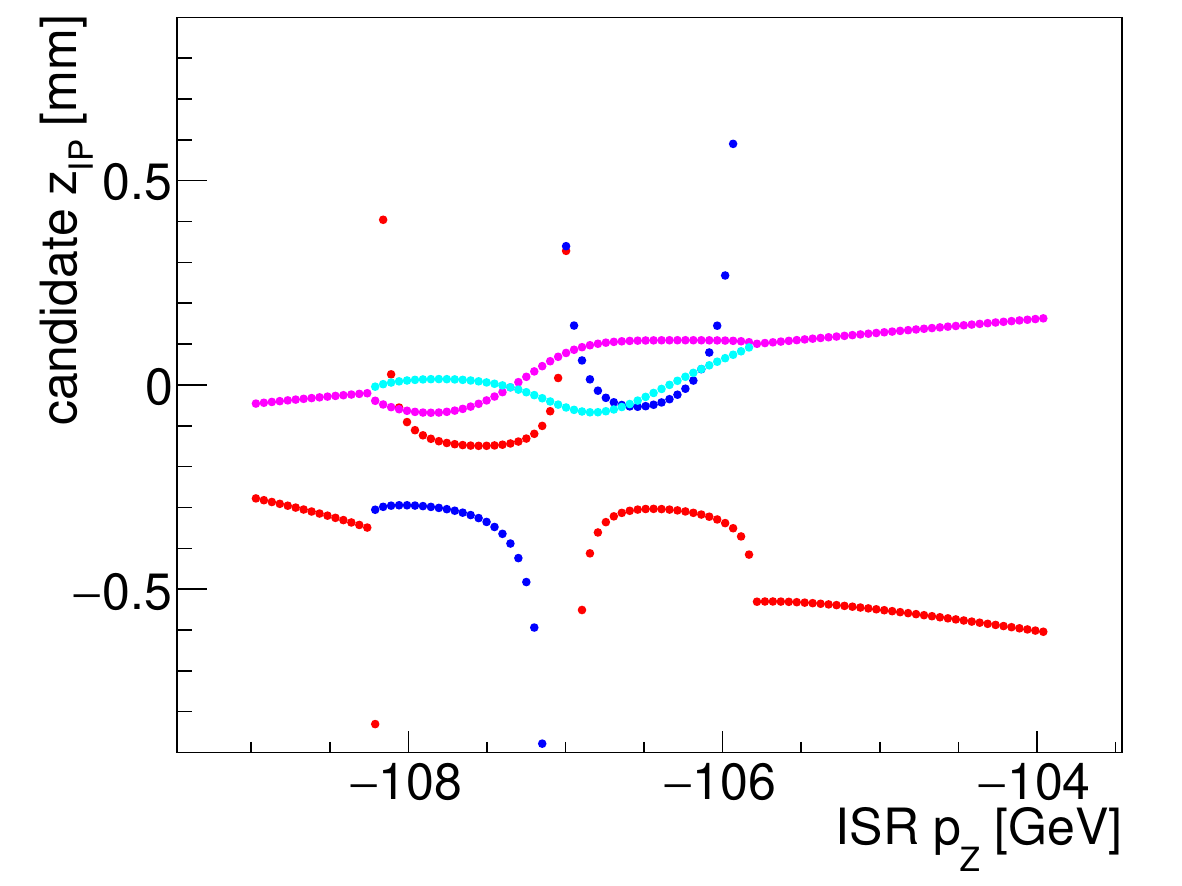}
\caption{Example events reconstructed assuming perfect detector response
  and assuming a single unseen ISR photon.
  Plots in the upper (lower) half show events with small (large) ISR.
  The extracted tau production position along the $z$ axis, $z_{IP}$, is shown for different assumed momenta $p_Z$ of the ISR photon.
  There are up to two event solutions per assumed $p_Z$: the first is shown in (red/magenta) and the second in (blue/cyan) for the $\tau^{(-/+)}$.
  Good solutions occur when the red \& magenta or blue \& cyan lines are consistent within experimental uncertainties, and
  $z_{IP}$ is consistent with the luminous region (approximately gaussian with width $\sim 200 \mu m$ at ILC-250).
}
\label{fig:examples}
\end{center}
\end{figure*}

In this section we assume a perfect measurement of charged pions and photons, and apply the reconstruction method
described in the previous section.
To illustrate the method, Fig.~\ref{fig:examples} shows the result of single ISR photon
$p_{ISR}$ scans in a selection of events with $m_{\tau\tau} \sim 250~\gev$ and $\sim 91~\gev$.
The displayed range in the ISR photon's momentum $p_{ISR}$ is that in which real tau cone solutions intersect.
The curves show the value of $z_{IP}$ calculated for each tau in the two solutions at each assumed $p_{ISR}$.
Intersections between these curves represent points at which the impact parameters are consistent with
the kinematics.
Given the complex forms of the curves, it unfortunately seems impossible to extract these points analytically.

\begin{figure}
  \begin{center}
    \includegraphics[width=0.3\textwidth]{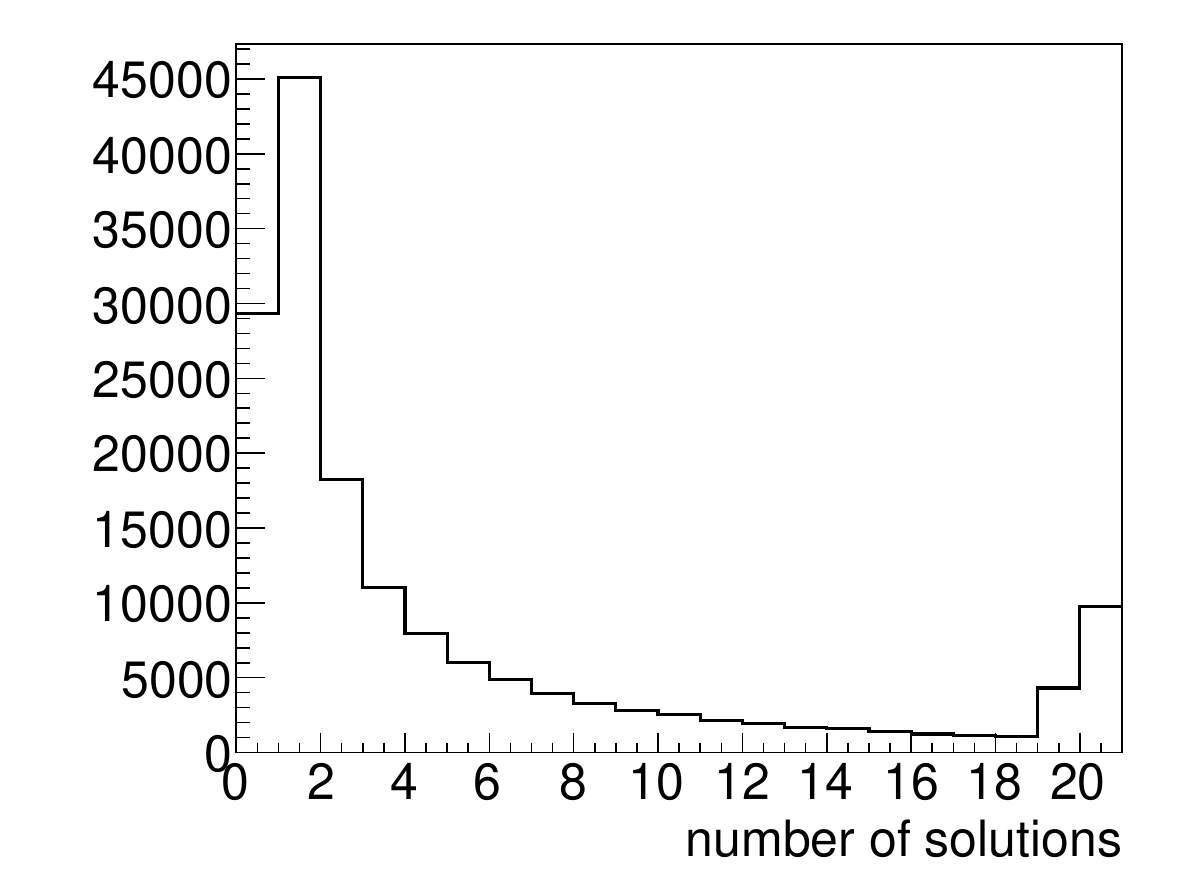}
    \caption{Number of solutions selected per event.
      Perfect detector response is assumed.
    }
  \end{center}
  \label{fig:nsol}
\end{figure}

\begin{figure*}
  \begin{center}
    \includegraphics[width=0.3\textwidth]{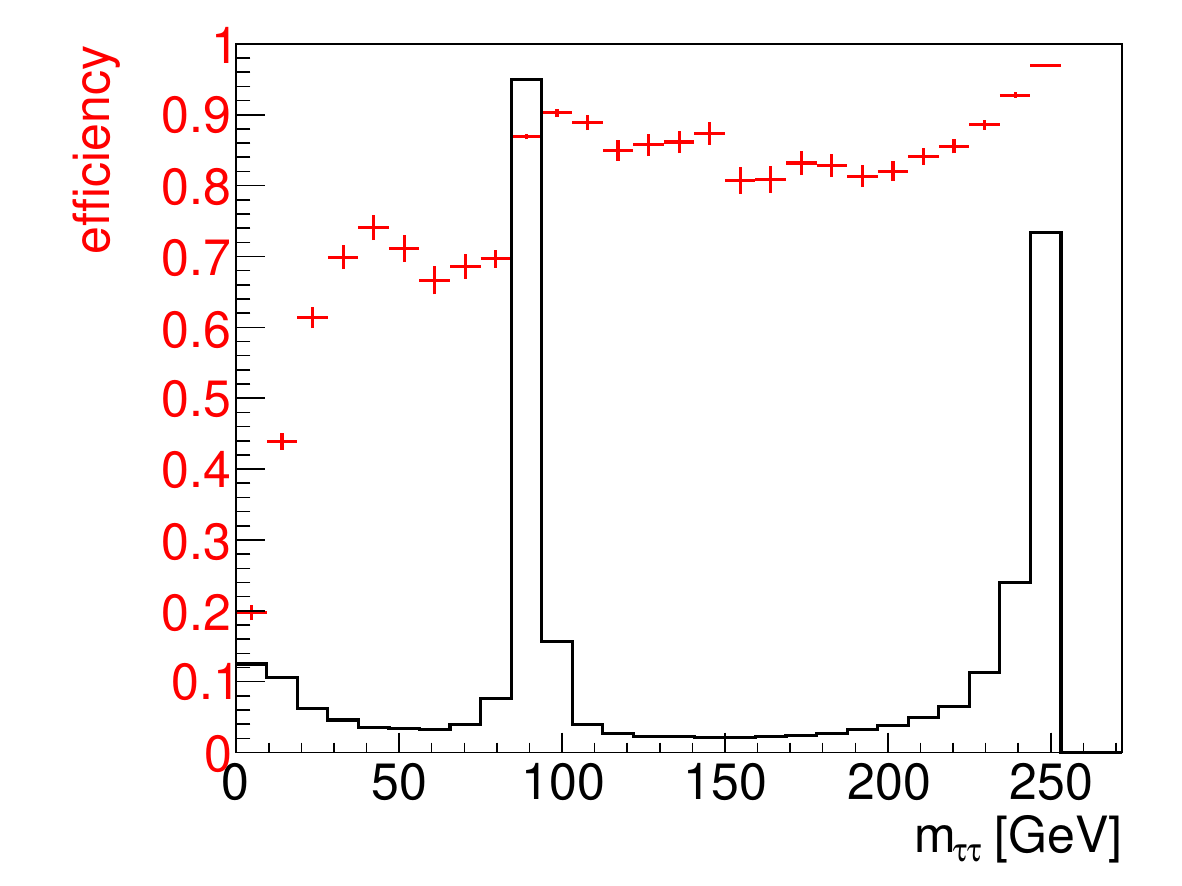} 
    \includegraphics[width=0.3\textwidth]{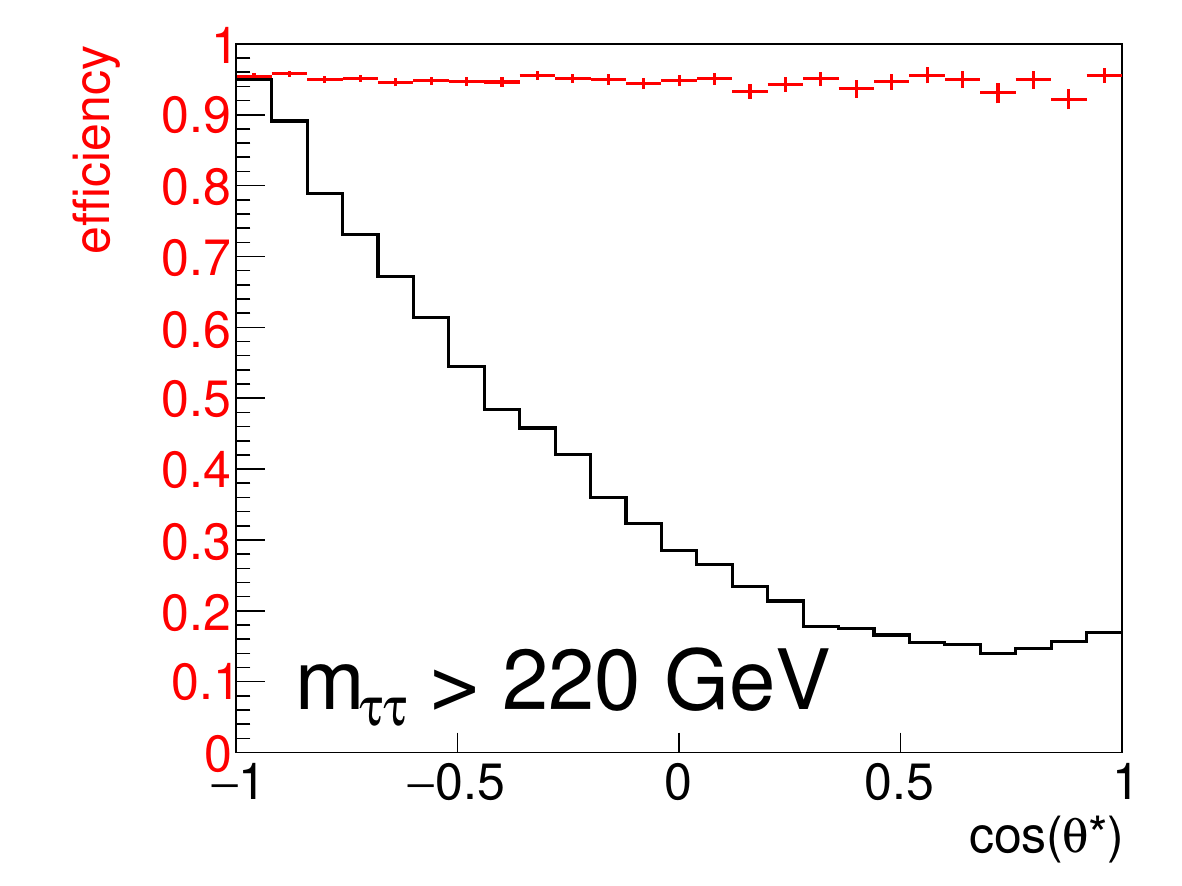}
    \includegraphics[width=0.3\textwidth]{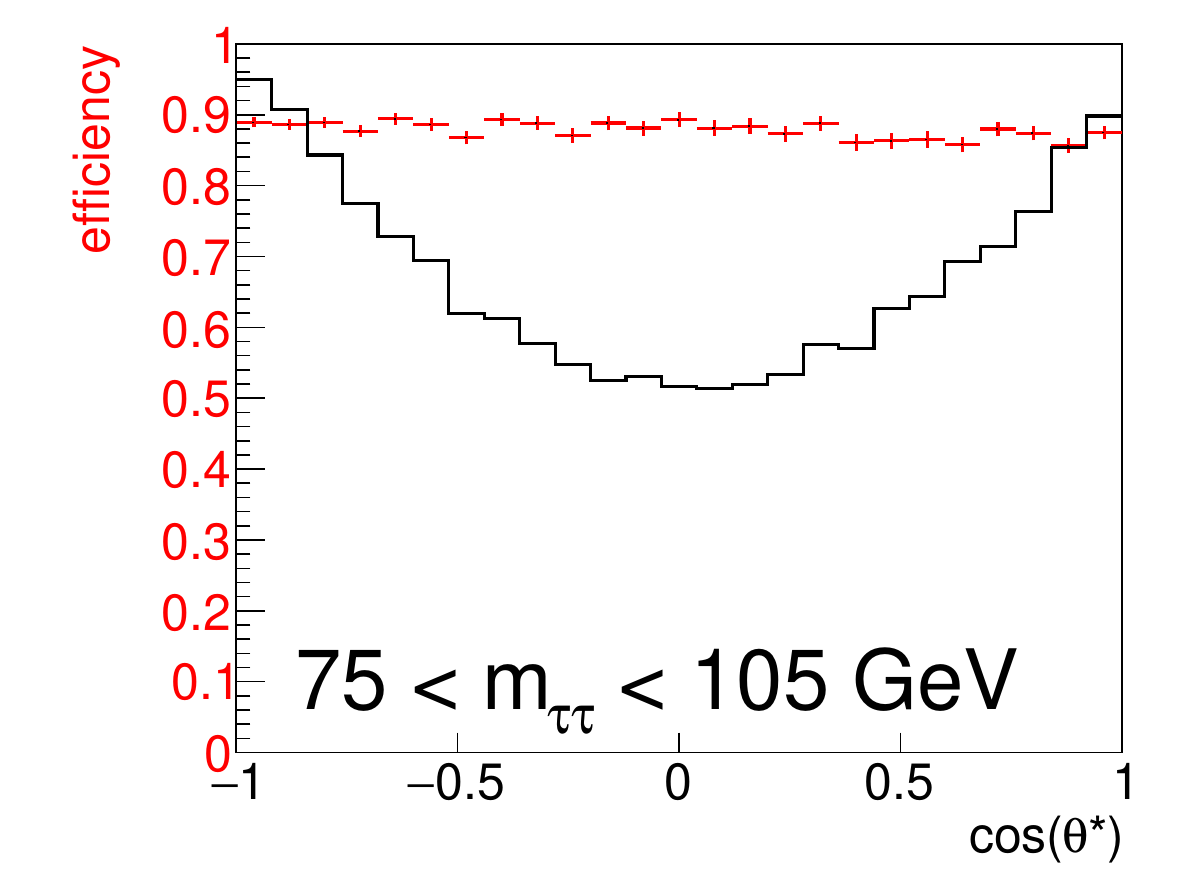}
    \caption{
      Distribution of events (black), and efficiency to find at least one solution (red) as a function of the
      true tau-tau invariant mass (left),
      and the CoM scattering angle for high-- (middle) and close-to-Z--mass (right) events.
      Perfect detector response is assumed.
    }
  \end{center}
  \label{fig:effs}
\end{figure*}

Figure~\ref{fig:nsol} shows the number of chosen solutions per event, and Fig.~\ref{fig:effs} the efficiency
to find at least a single good solution as a function of the tau-tau invariant mass
and the CoM scattering angle, both for events at high mass and around the Z peak.
The efficiency is well above 95\% close to the kinematic limit at 250~\gev,
drops to around 90\% at the Z pole, and falls further at smaller masses.
No significant dependence of the efficiency on the scattering angle is seen.

\begin{figure*}
  \begin{center}
    \includegraphics[width=0.3\textwidth]{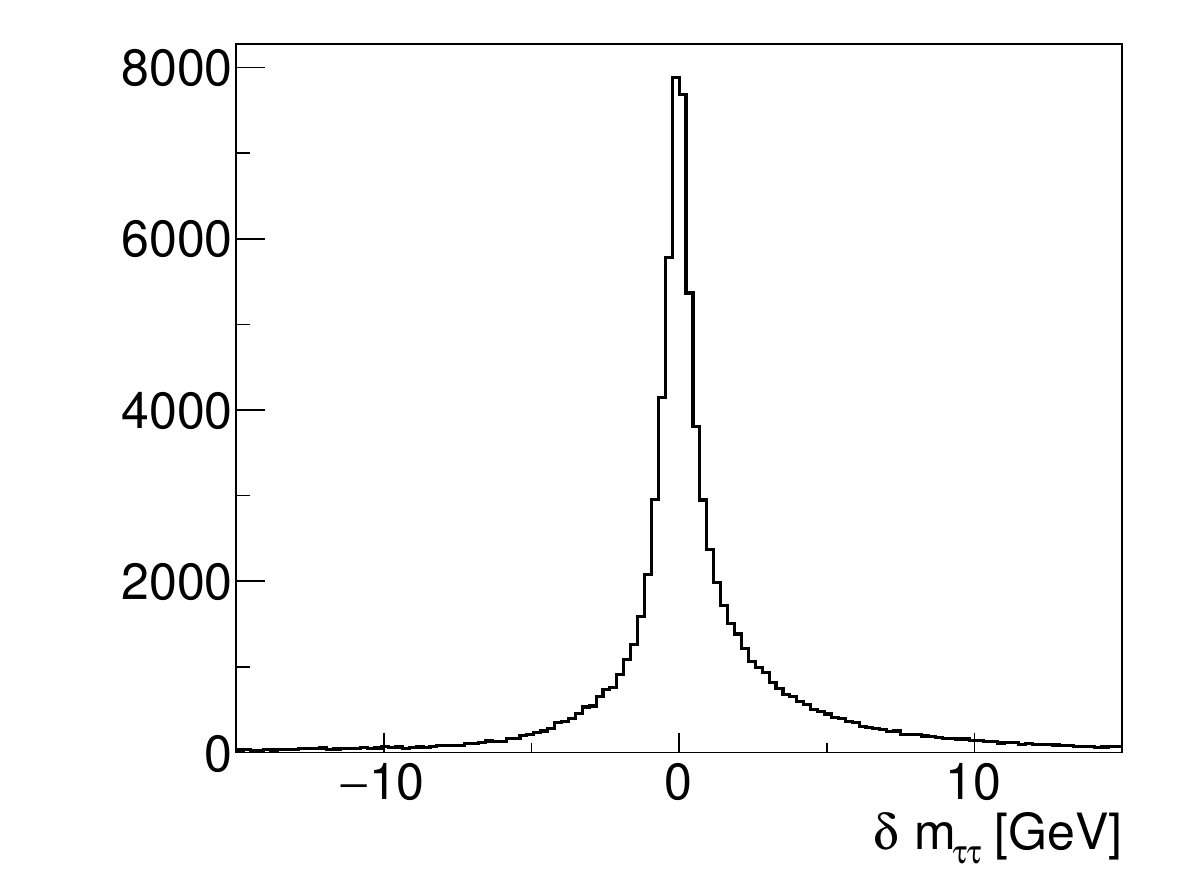} 
    \includegraphics[width=0.3\textwidth]{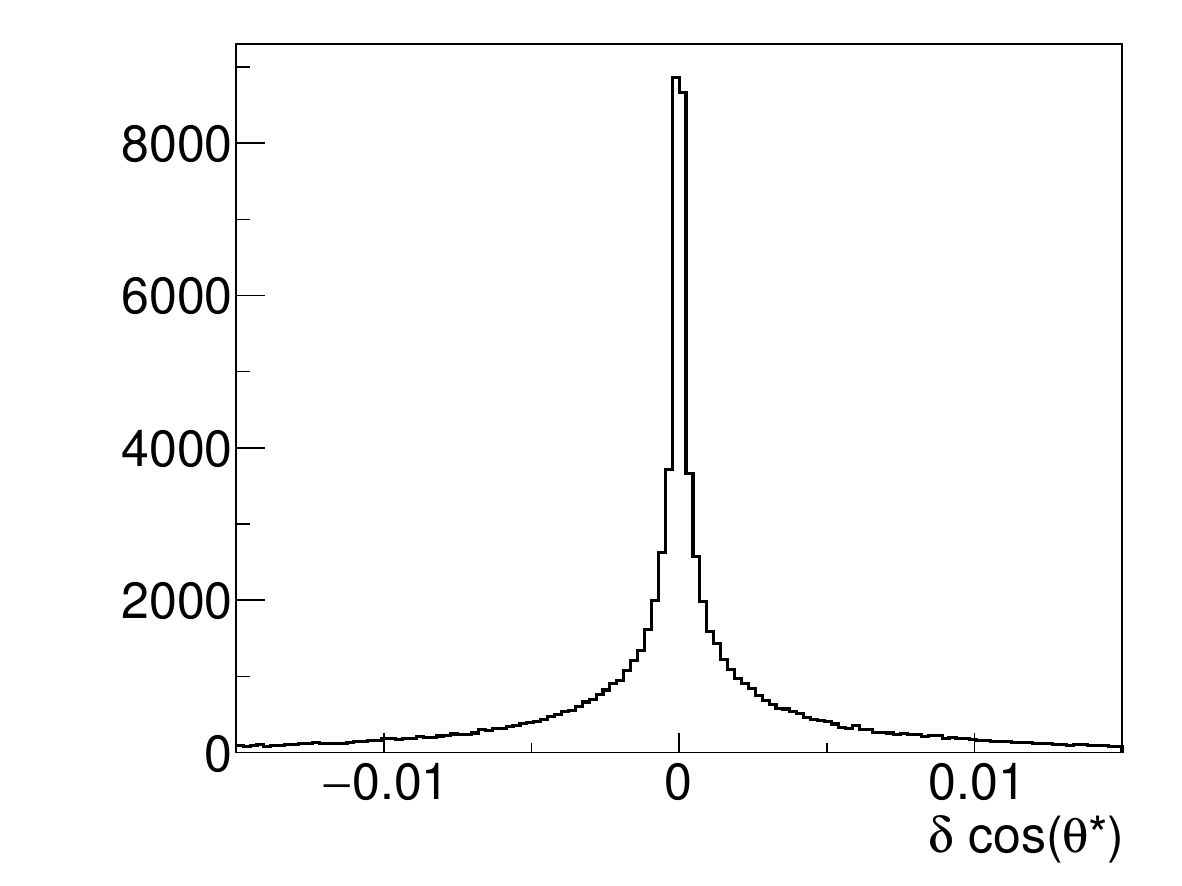}
    \includegraphics[width=0.3\textwidth]{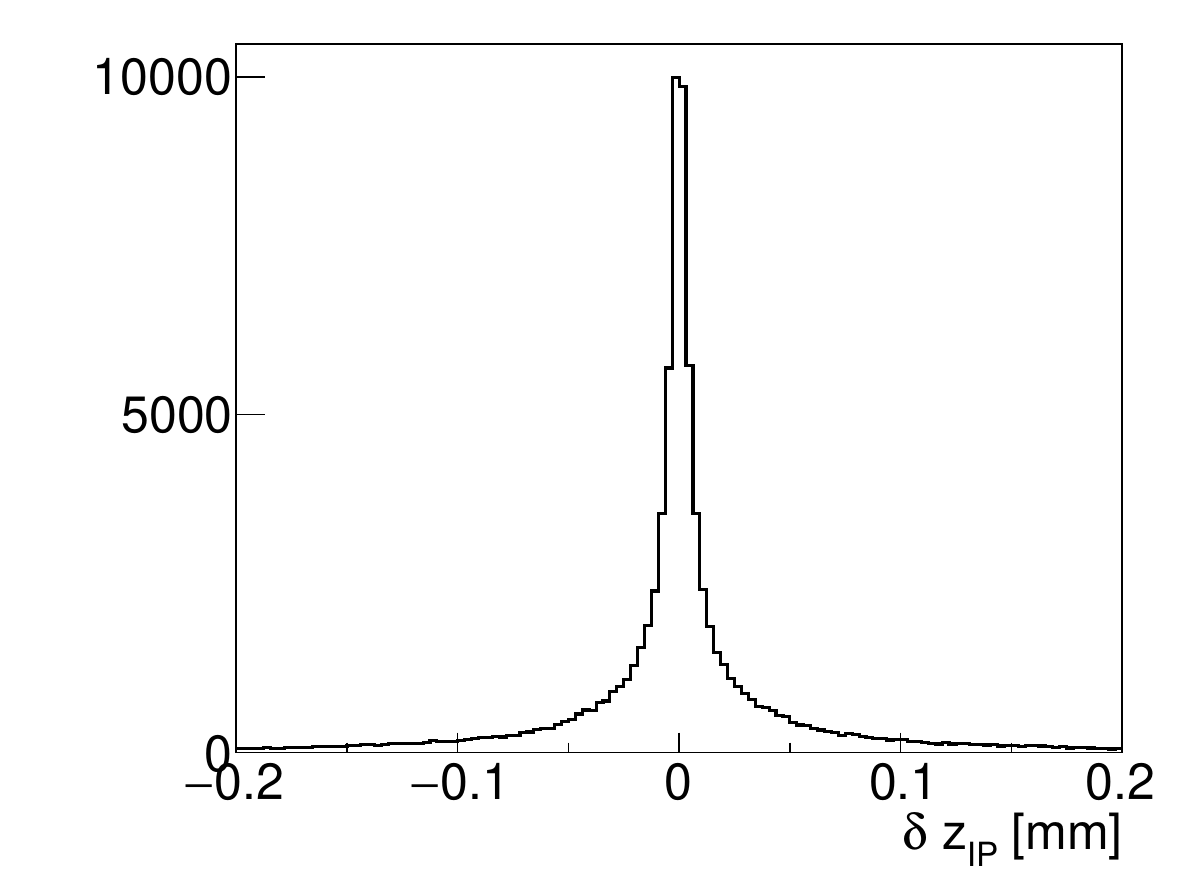}
    \caption{The difference between the reconstructed and true values of the tau-tau invariant mass (left), the CoM scattering angle
      $\cos \theta^*$ (center), and the position of the IP $z_{IP}$ (right).
      Perfect detector response assumed.
    }
  \end{center}
  \label{fig:mcrel}
\end{figure*}

To address the precision with which the event kinematics are reconstructed, Fig.~\ref{fig:mcrel}
shows the difference between the reconstructed and true values of $m_{\tau\tau}$,
the cosine of the CoM scattering angle $\cos \theta^*$, and the IP position $z_{IP}$.
The resolution on $m_{\tau\tau}$ is around a \gev\ (but with some non-gaussian tails),
and on $\cos \theta^*$ is $\mathcal{O}(0.001)$.
The position of the IP $z_{IP}$ is reconstructed to $\mathcal{O}(10) \mu m$.

\section{Polarimeter extraction}

\label{sec:spin}

Since we have fully reconstructed the kinematics of a pair of hadronically decaying taus, we can extract
polarimeters with optimal sensitivity to the their spin orientation.
We use the momenta of the decay products in the rest frame of the mother tau.
Each decay mode has a specific optimal polarimeter, a function of the decay daughters' momenta.

In the case of 1- and 2-pion decays, we define the polarimeter vector $\textbf{h}$ as
\begin{eqnarray}
\mathbf{h} ( \taupi )  \propto && \mathbf{p}_{\pi^\pm} \\
\mathbf{h} ( \taurho ) \propto && 2 ( q \cdot p_\nu ) \mathbf{q}  - q^2 \mathbf{p}_\nu ,
\label{eqn:polar}
\end{eqnarray}
where $p_{\pi^\pm}$, $p_{\pi^0}$, $p_\nu$ are respectively the four-momenta of the charged and neutral pions,
and of the neutrino defined in the tau rest frame, and $q = p_{\pi^\pm} - p_{\pi^0}$~\cite{Jadach:1990mz}.
For the more complex three-pion decays $\tau^\pm \to \pi^\pm\pi^0\pi^0\nu$ and $\tau^\pm \to \pi^\pm\pi^\pm\pi^\mp\nu$
we use the method and associated computer code of~\cite{Cherepanov:2023wfp}.

\begin{figure*}
  \begin{center}
    \includegraphics[width=0.24\textwidth]{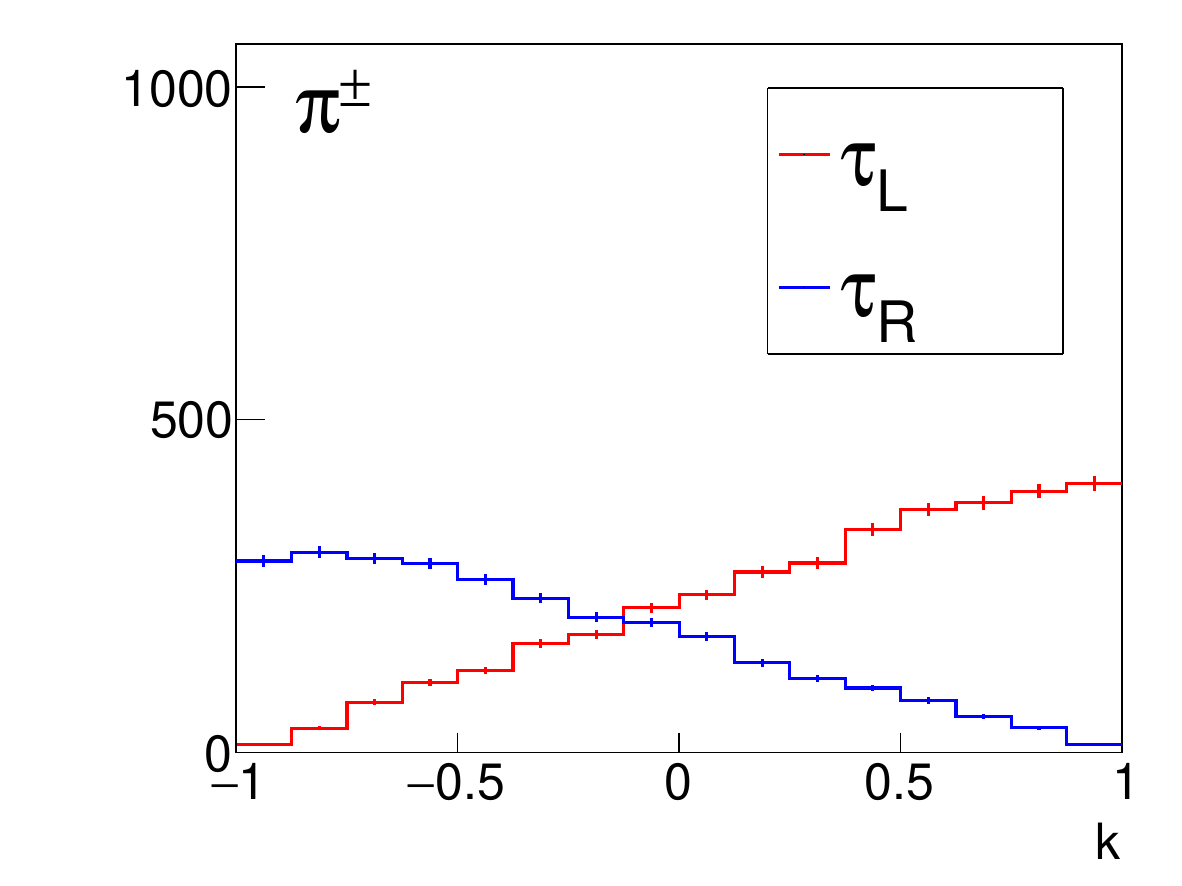}
    \includegraphics[width=0.24\textwidth]{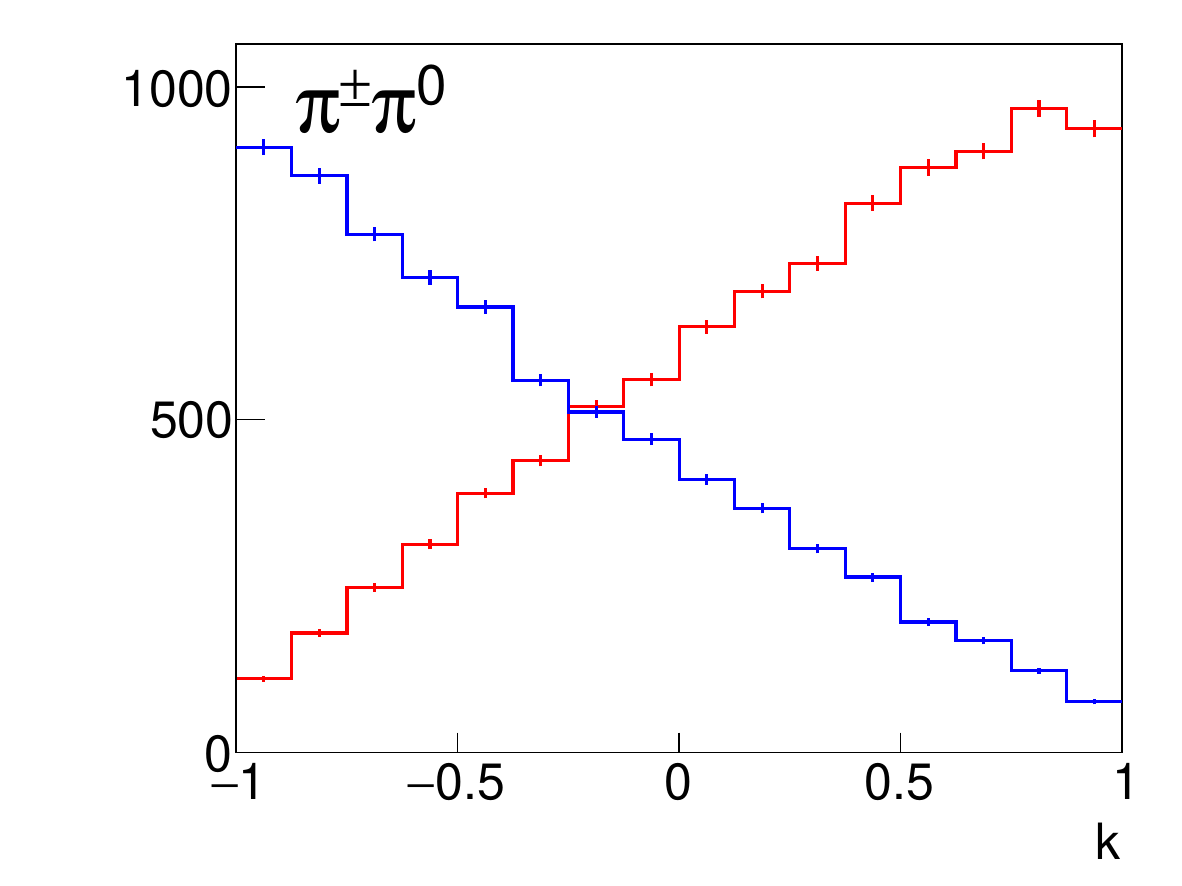}
    \includegraphics[width=0.24\textwidth]{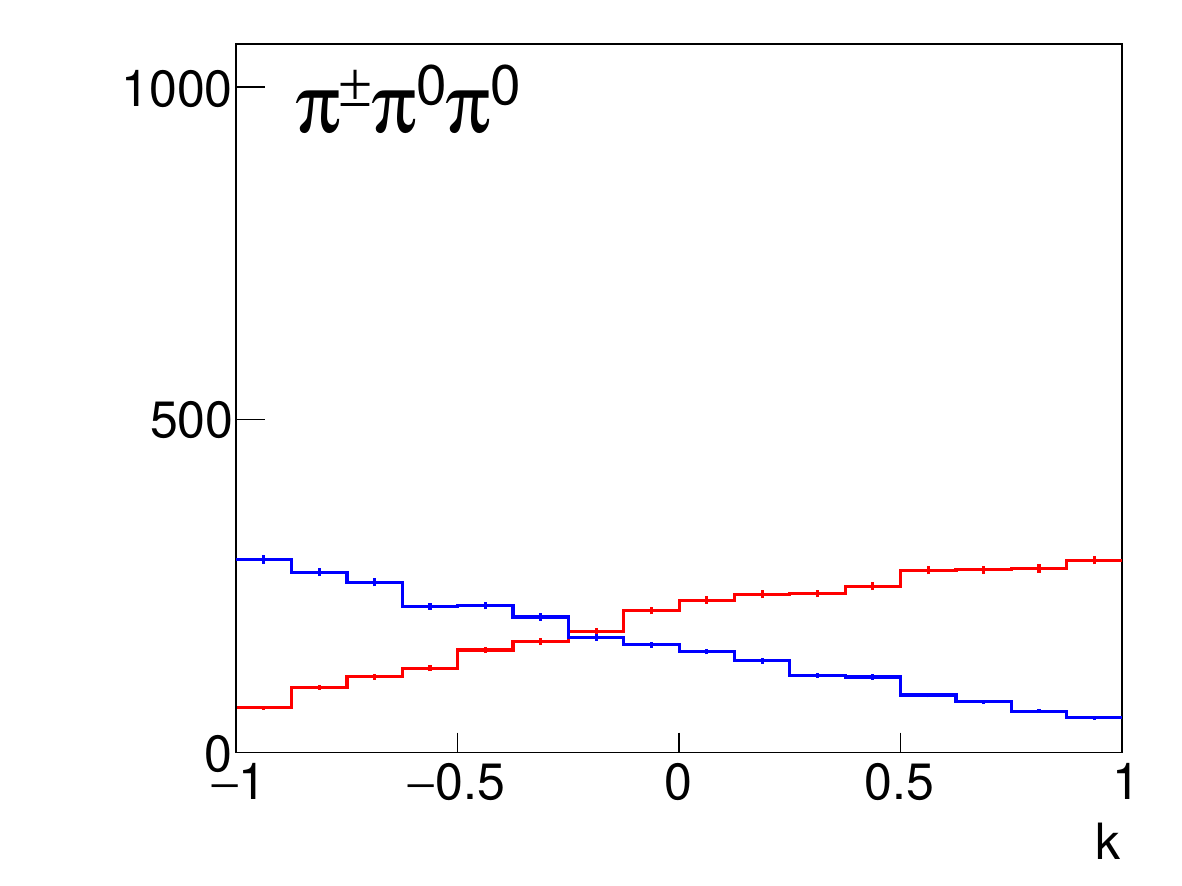}
    \includegraphics[width=0.24\textwidth]{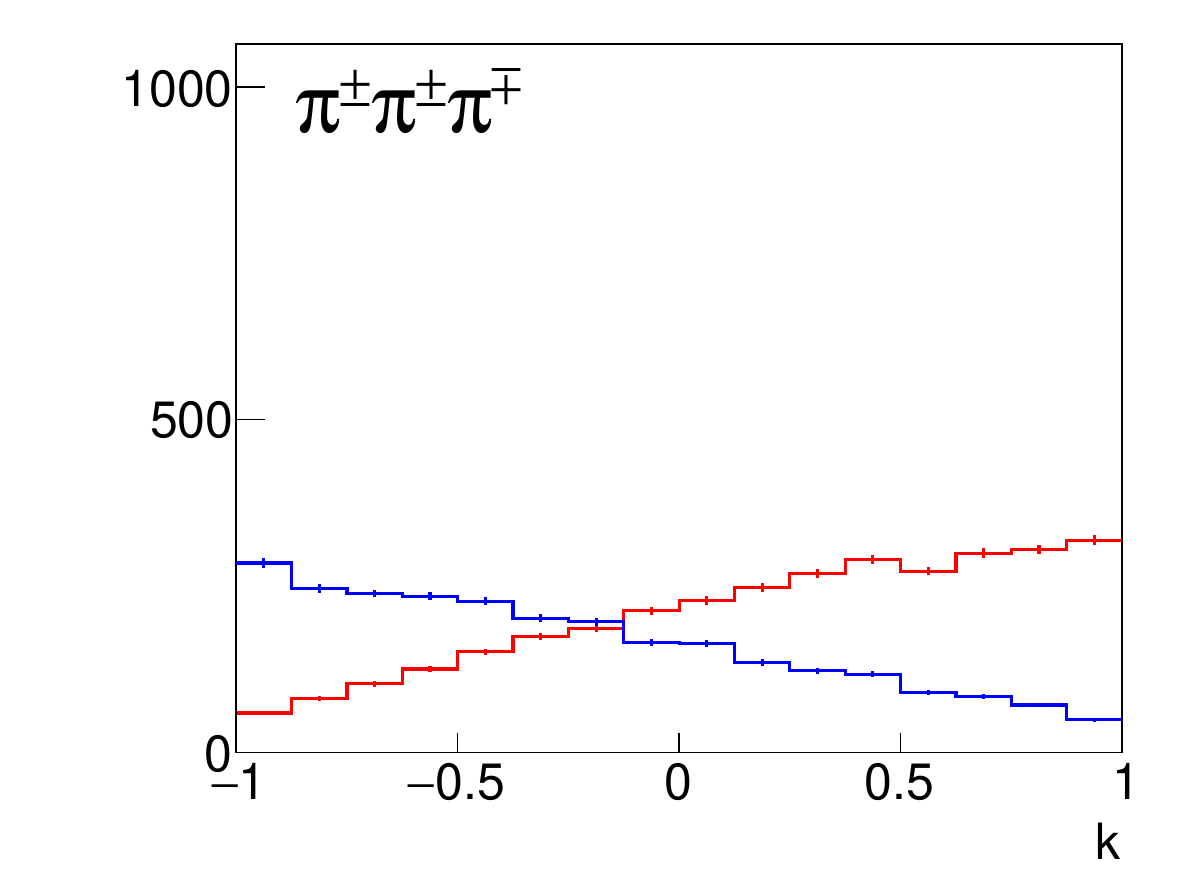}
    \caption{$k$ projection of reconstructed polarimeters for $\tau_{L/R}$ decays to 1, 2, and 3 pions for events with $m_{\tau\tau} > 220~\gev$,
      assuming perfect detector response.}
    \label{fig:pols}
  \end{center}
\end{figure*}

The longitudinal polarisation of the tau can be extracted from the projection of the polarimeter
vector onto the tau momentum direction.
This is shown in Fig.~\ref{fig:pols} separately for the four decay channels, and for left and right-handed taus.
The reconstructed distributions are close to the ideal distributions $f_{L/R}(k) \propto 1\ \sfrac{+}{-} \ k$,
with only modest dilution due to the possible presence of multiple incorrect solutions per event.

\begin{figure*}
  \begin{center}
    \includegraphics[width=0.75\textwidth]{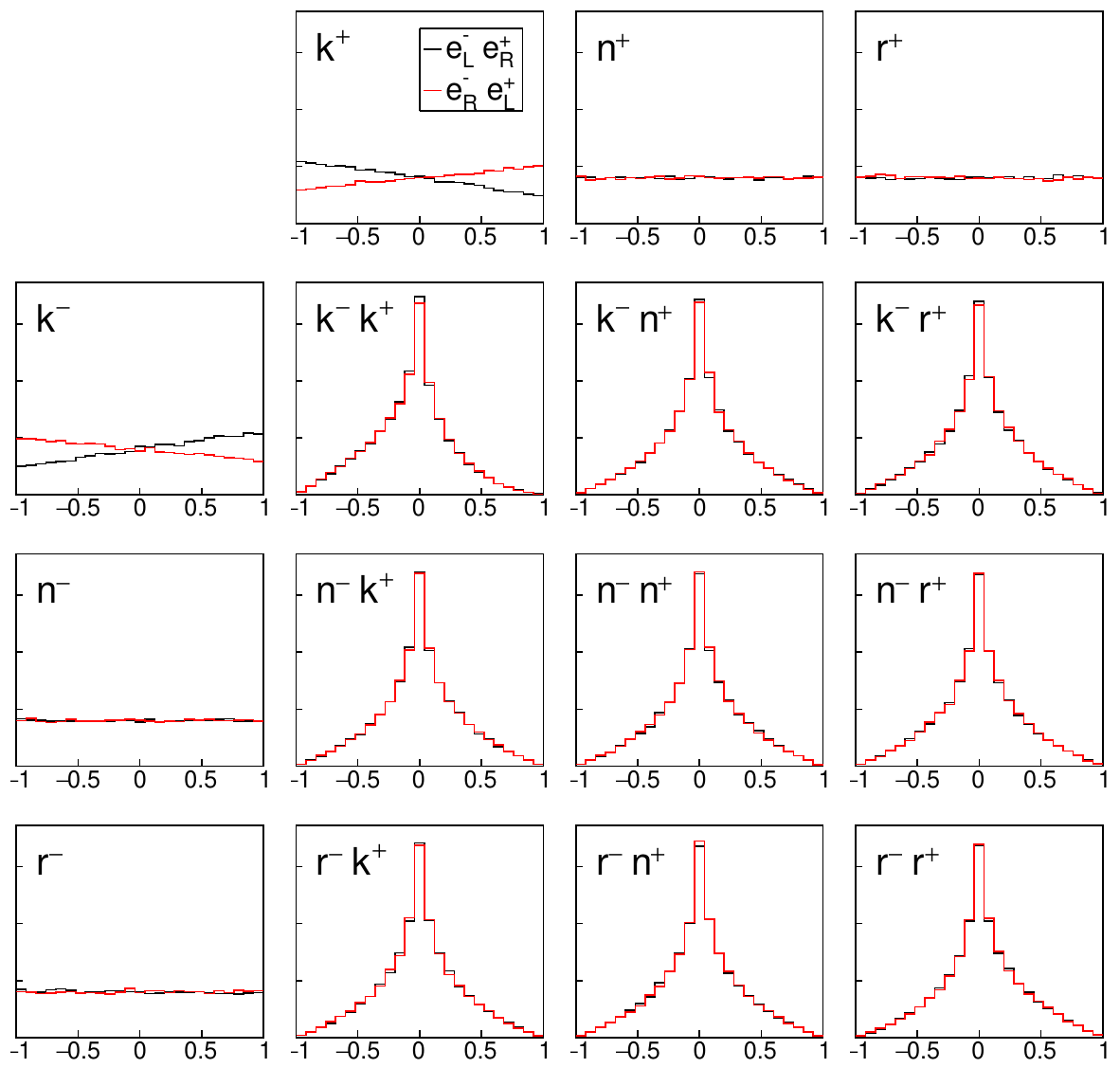}
    \caption{Distributions of the reconstructed spin density matrix elements in the helicity basis ($k, n, r$), summing over all $m_{\tau\tau}$,
      calculated using the true momenta of taus and all decay products.
      The predictions for 100\% polarised initial beams $e^-_L e^+_R$ and $e^-_R e^+_L$ are compared.
    }
    \label{fig:spinDensMC}
  \end{center}
\end{figure*}

Full tomography of the tau's quantum state requires us to also consider the other projections of the polarimeter, as well
as the correlations between the two taus.
We use the helicity basis coordinates~\cite{Baumgart:2012ay} within the tau-tau rest frame: $k$ is along the $\tau^-$ momentum
direction,
$n$ is perpendicular to the plane defined by $k$ and the beam momenta, and $r$ is with the beam-tau plane and perpendicular to $k$.
Figure~\ref{fig:spinDensMC} shows the projections of the reconstructed polarimeter onto the three axes
separately for the two taus ${k^\pm, n^\pm, r^\pm}$, as well as the products of these components between the
two taus, for different initial beam polarisation states. This set of observables fully characterises the
spin state of the tau pair, and therefore allows the computation of various measures (such as the purity, concurrence, magic, Bell-CHSH)
sensitive to quantum entanglement.

\begin{figure*}
  \begin{center}
    \includegraphics[width=0.75\textwidth]{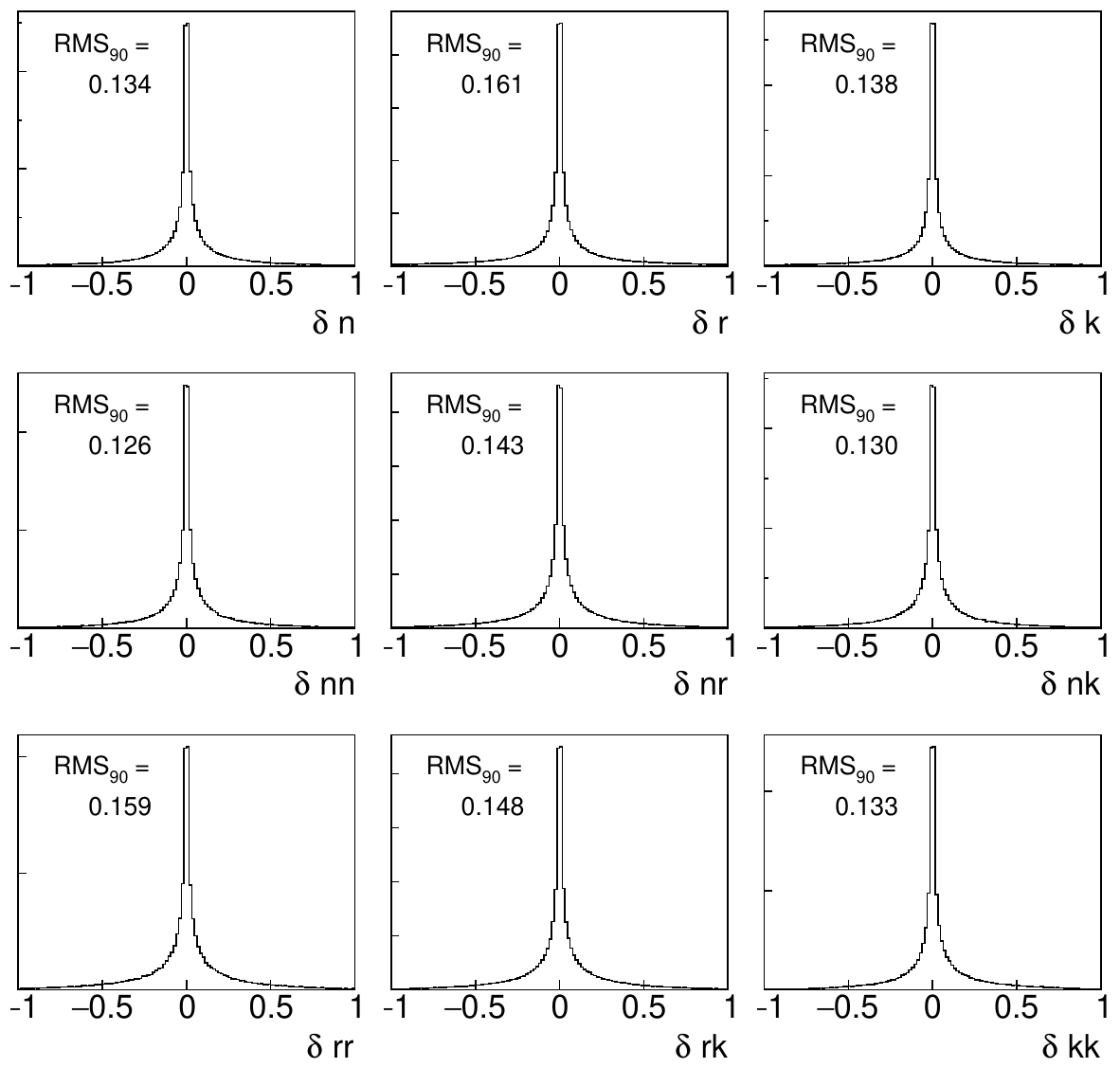}
    \caption{Distributions of the difference between true and reconstructed spin density matrix elements and their
      RMS90 measure, assuming a detector with perfect response.
    }
  \end{center}
  \label{fig:mcrelSp}
\end{figure*}

Figure~\ref{fig:mcrelSp} compares the spin density elements extracted from the reconstructed polarimeters
assuming perfect detector response, 
with those calculated using the true momenta of all tau decay products (including the neutrino).
The resolution on each of these quantities is not gaussian; the central peak is much narrower than 0.1,
with relatively large tails due to the presence of incorrect solutions.
The RMS90 of the distributions (the RMS of the population which lies within the smallest range which contains at least 90\% of entries)
is around 0.15 for all components.
Since this resolution is significantly smaller than the range [-1,1] of these observables,
it implies that a precise quantum tomography of this process is in principle possible,
at least given a detector with perfect response.

\section{Effects of detector resolution}

\label{sec:detres}

To investigate the effects of detector resolution, we use the SGV fast simulation package~\cite{Berggren:2012ar}
to simulate the performance of the International Large Detector (ILD) concept~\cite{ILDConceptGroup:2020sfq}.
The ILD consists of a precise silicon-based vertex detector, a large hybrid tracking system consisting of a Time Projection Chamber and silicon detector layers,
and highly granular calorimeters optimised for particle flow reconstruction.
The key detector parameters for this analysis are the material budget and single hit resolution of the vertex detector
which determine how well the impact parameter of tracks is measured,
and the energy and angular resolution of the electromagnetic calorimeter which determine the measurement of photon momenta.
The momentum resolution of the tracker used to reconstruct the charged pions is not expected to play a major role,
since it is significantly better than both the typical calorimetric photon energy resolution and
the intrinsic beam energy spread.

The default description of ILD in SGV assumes a vertex detector with three double layers,
with a single hit point resolution of $3 \mu m$ in each layer.
We varied this point resolution between 2 and 15~$\mu m$. The material budget in the inner region consists of the
beampipe with a thickness of $0.14\%~\mathrm{X_0}$ and each vertex detector layer's thickness of $0.11\%~X_0$.
We studied the effect of increasing this material by a factor of up to eight.

\begin{figure*} 
  \begin{center}
    \includegraphics[width=0.35\textwidth]{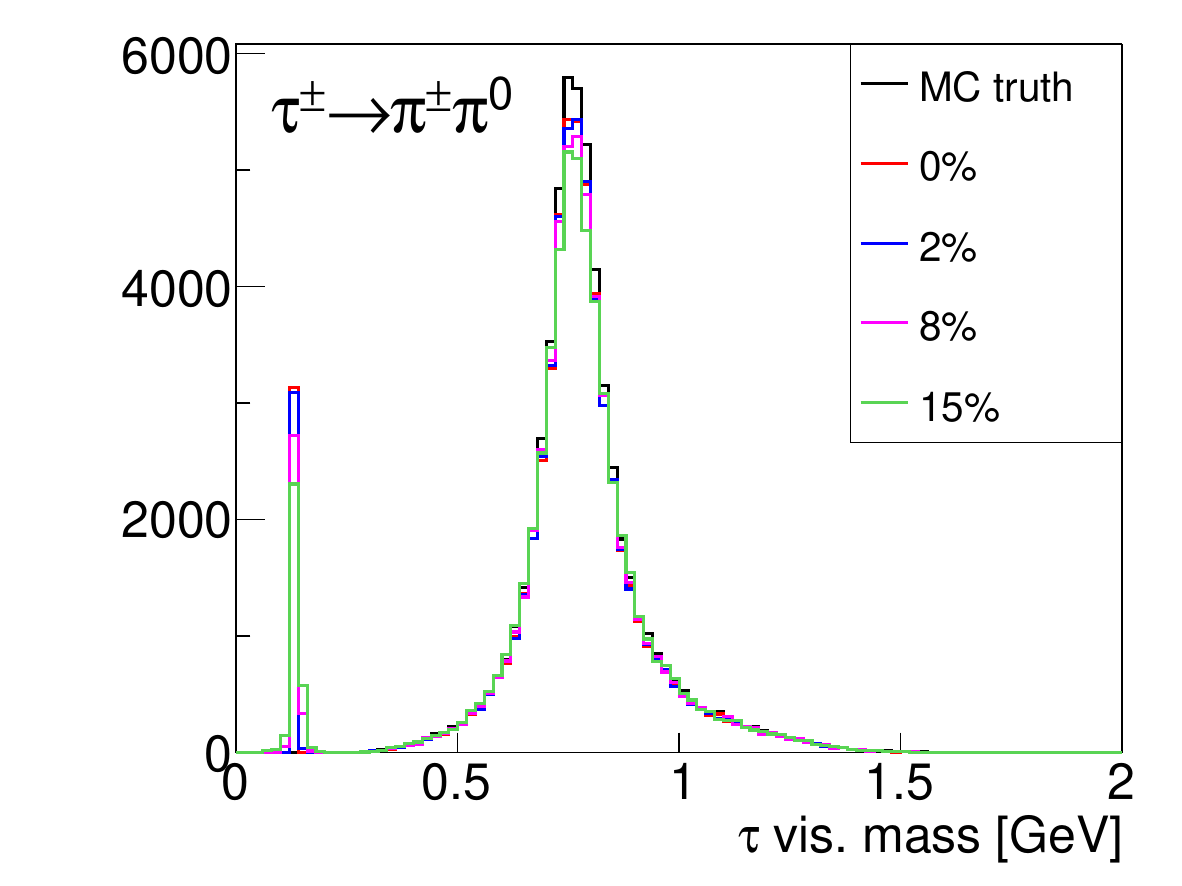}
    \includegraphics[width=0.35\textwidth]{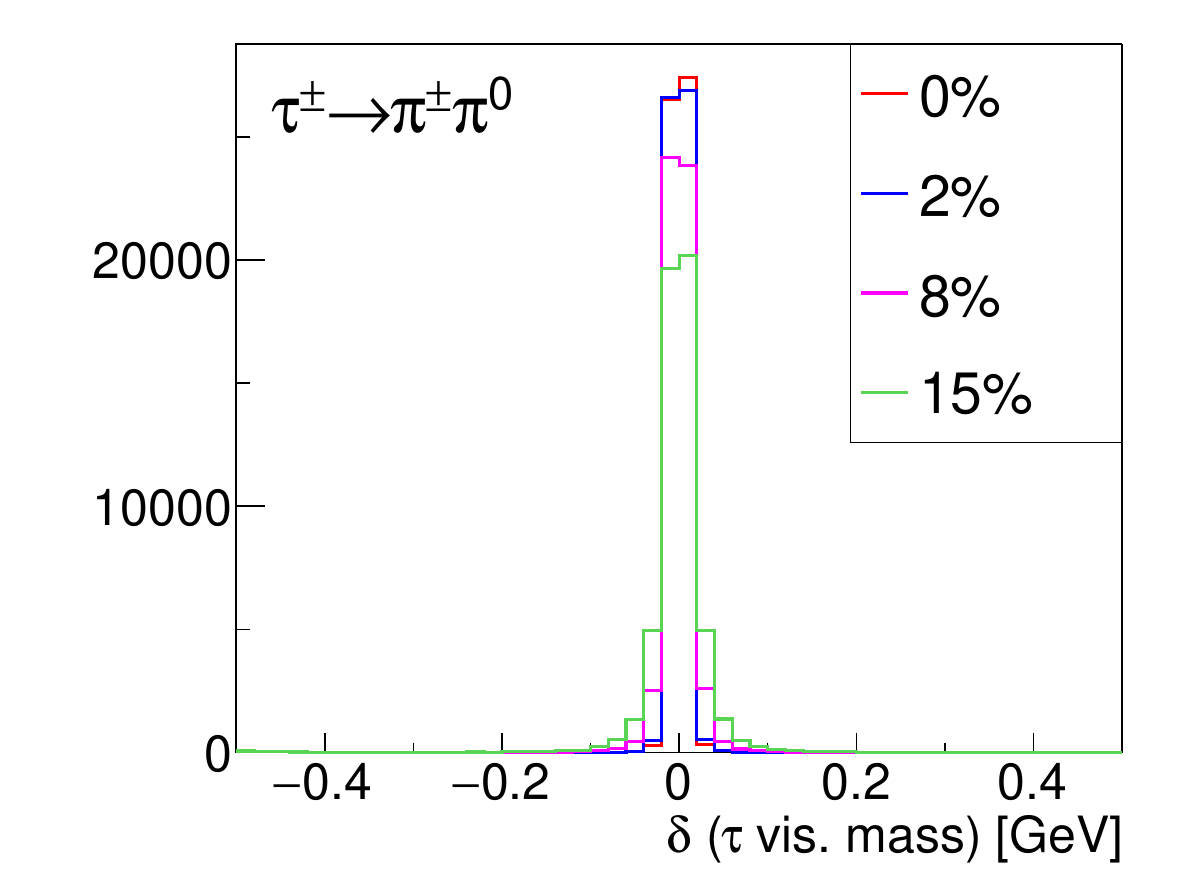} \\
    \includegraphics[width=0.35\textwidth]{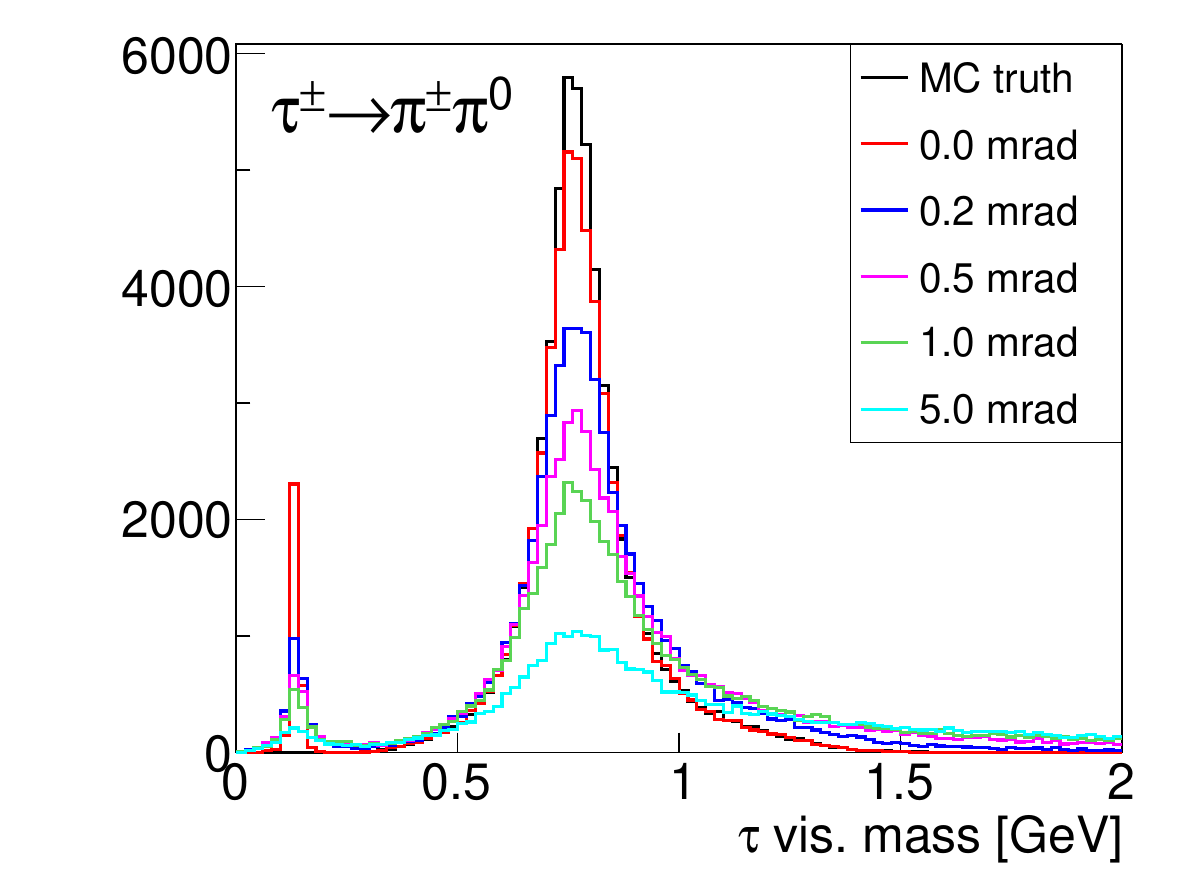}
    \includegraphics[width=0.35\textwidth]{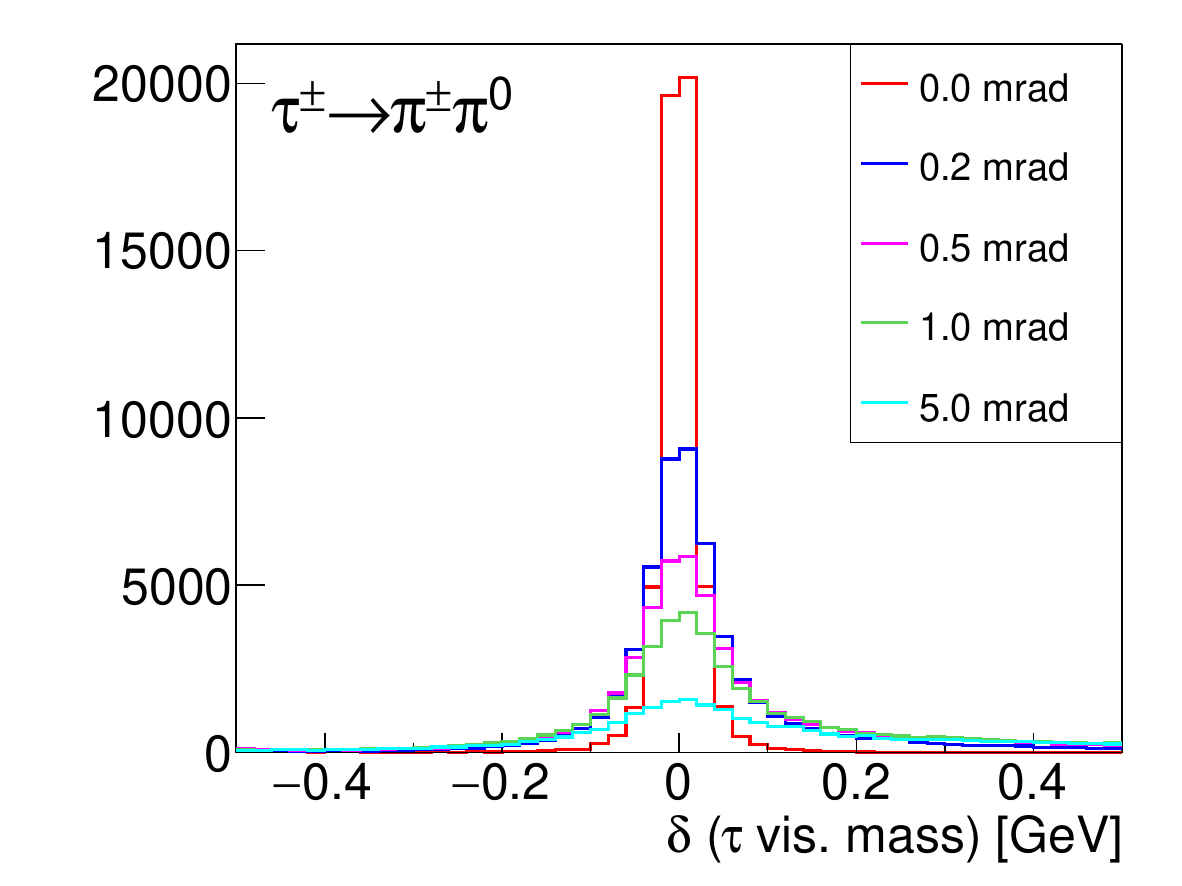}
    \caption{
      Tau decays to $\pi^\pm \pi^0$.
      The reconstructed visible invariant mass (left) and the
      event-by-event difference between the reconstructed and true visible mass (right)
      for ECALs of different energy (top) and angular (bottom) resolution. The population around
      the pion mass is due to cases in which a particle escapes the detector acceptance.
    }
    \label{fig:mvis}
  \end{center}
\end{figure*}

The energy resolution for photon measurement was modeled as the
quadratic sum of a stochastic term of (2, 8, or 15)\% combined with a constant term of 1\%.
The case with a 15\% stochastic term is representative of the highly-segmented sampling ECAL being considered for ILD.
An energy-independent photon direction smearing of (0, 0.2, 0.5, 1 or 5~mrad) was applied.
It was applied to both the azimuthal angle $\phi$ (scaled by a factor $1/\sin\theta$) and to the polar angle $\theta$,
to simulate an ideal ECAL with uniform cartesian segmentation at uniform distance from the IP.
The effect of these photon smearings on the reconstructed visible mass of the tau is shown in Fig.~\ref{fig:mvis}.
The energy smearing has only a rather small effect on the reconstructed mass,
while the smearing in angle has a much larger effect.

\begin{figure*}
  \begin{center}
    \includegraphics[width=0.35\textwidth]{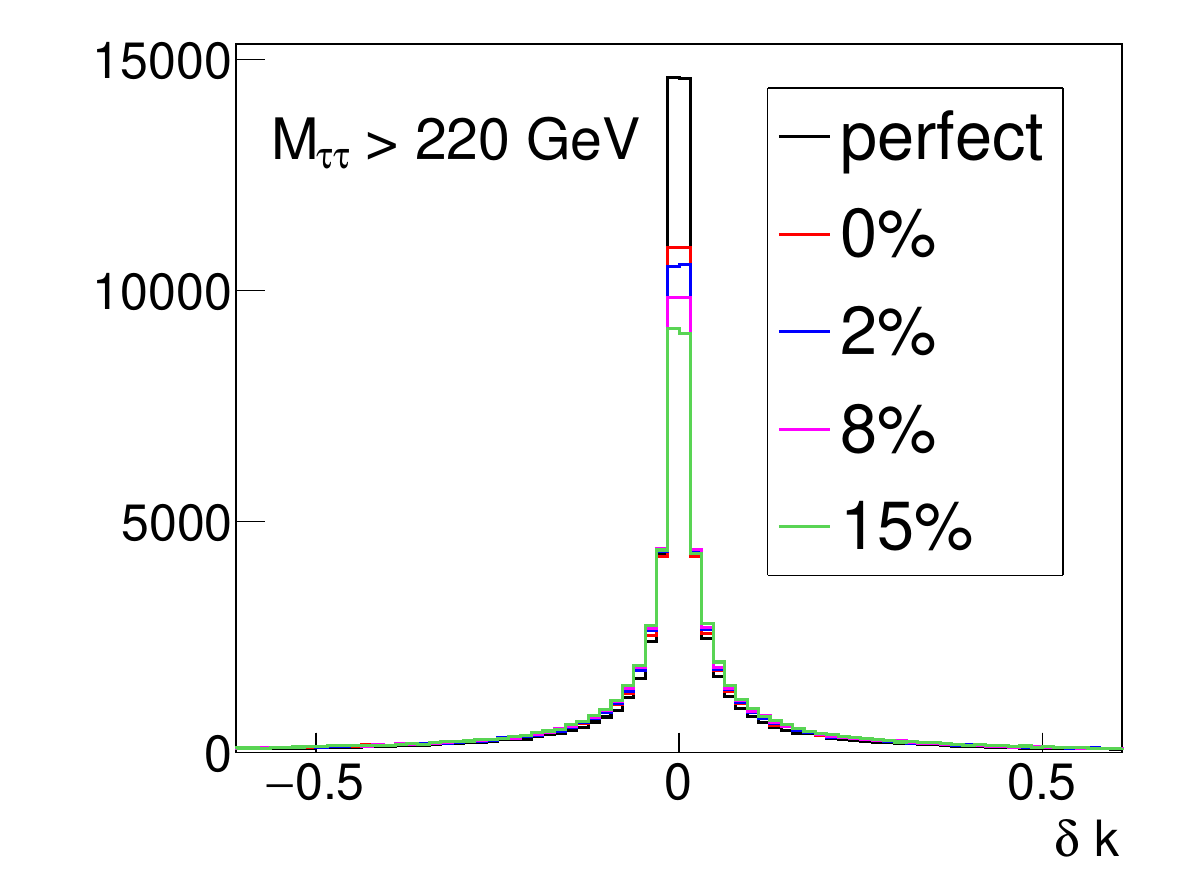}
    \includegraphics[width=0.35\textwidth]{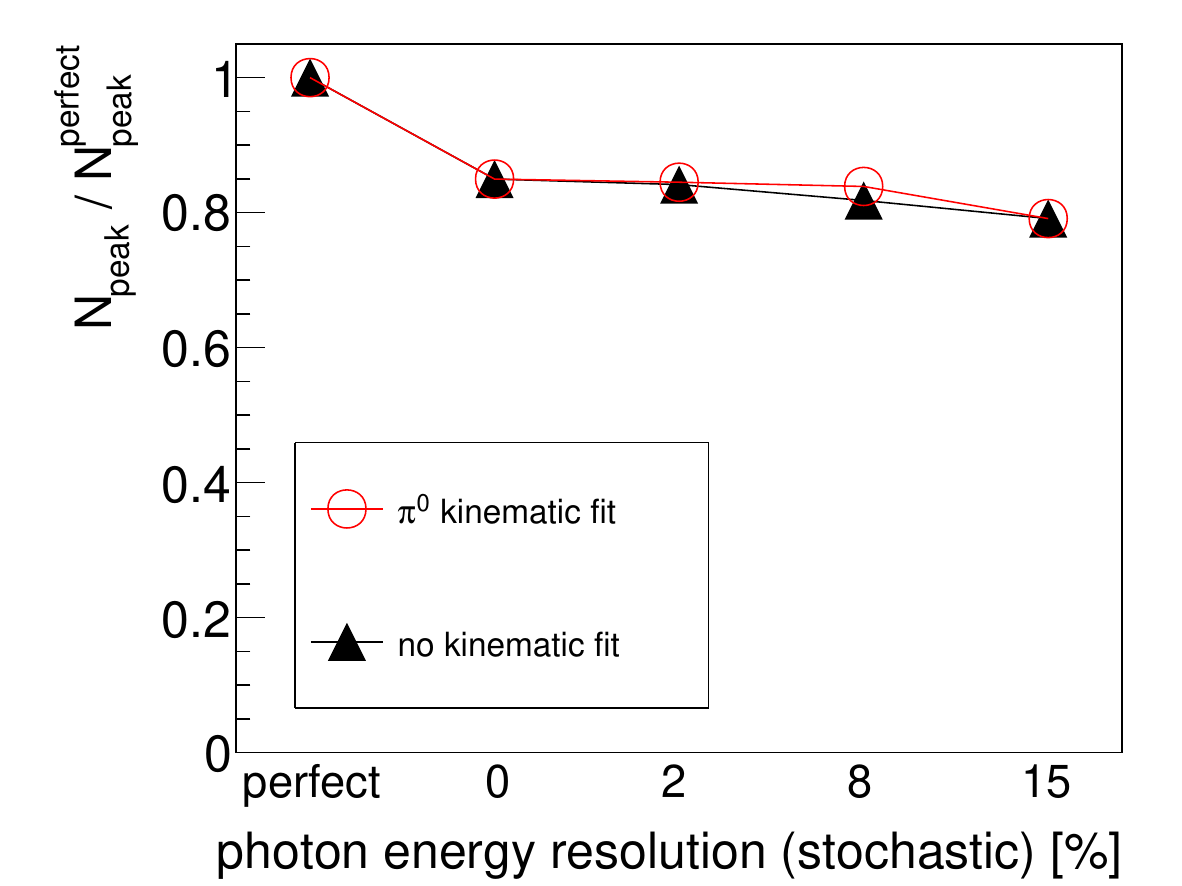} \\
    \includegraphics[width=0.35\textwidth]{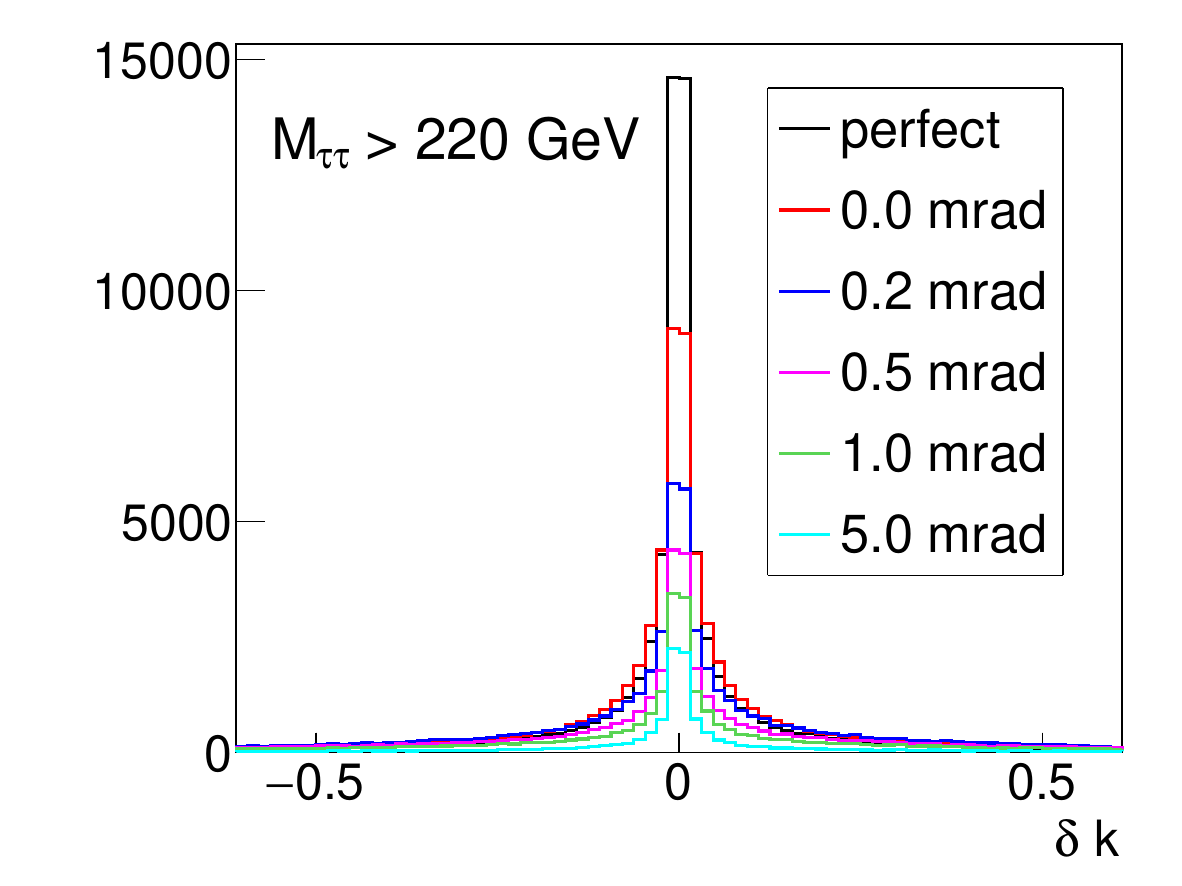}
    \includegraphics[width=0.35\textwidth]{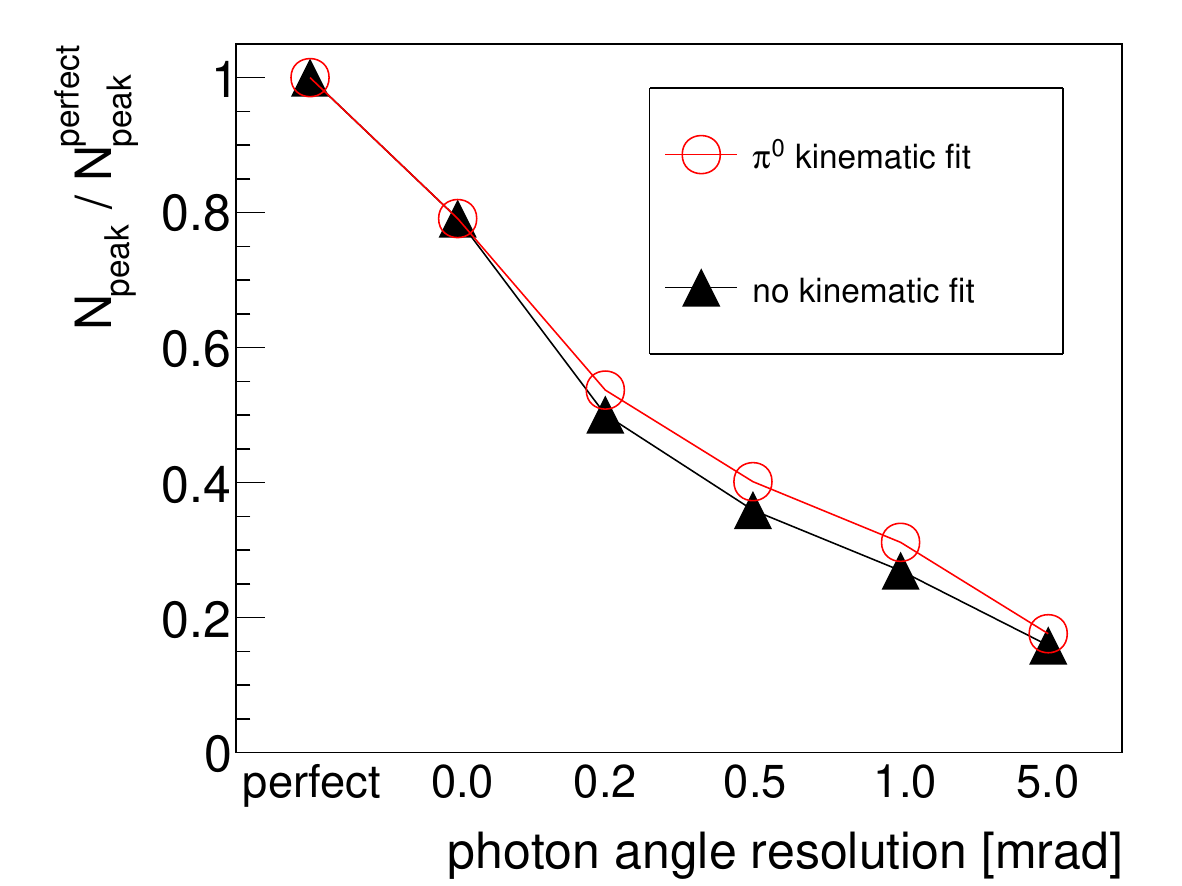} \\
    \includegraphics[width=0.35\textwidth]{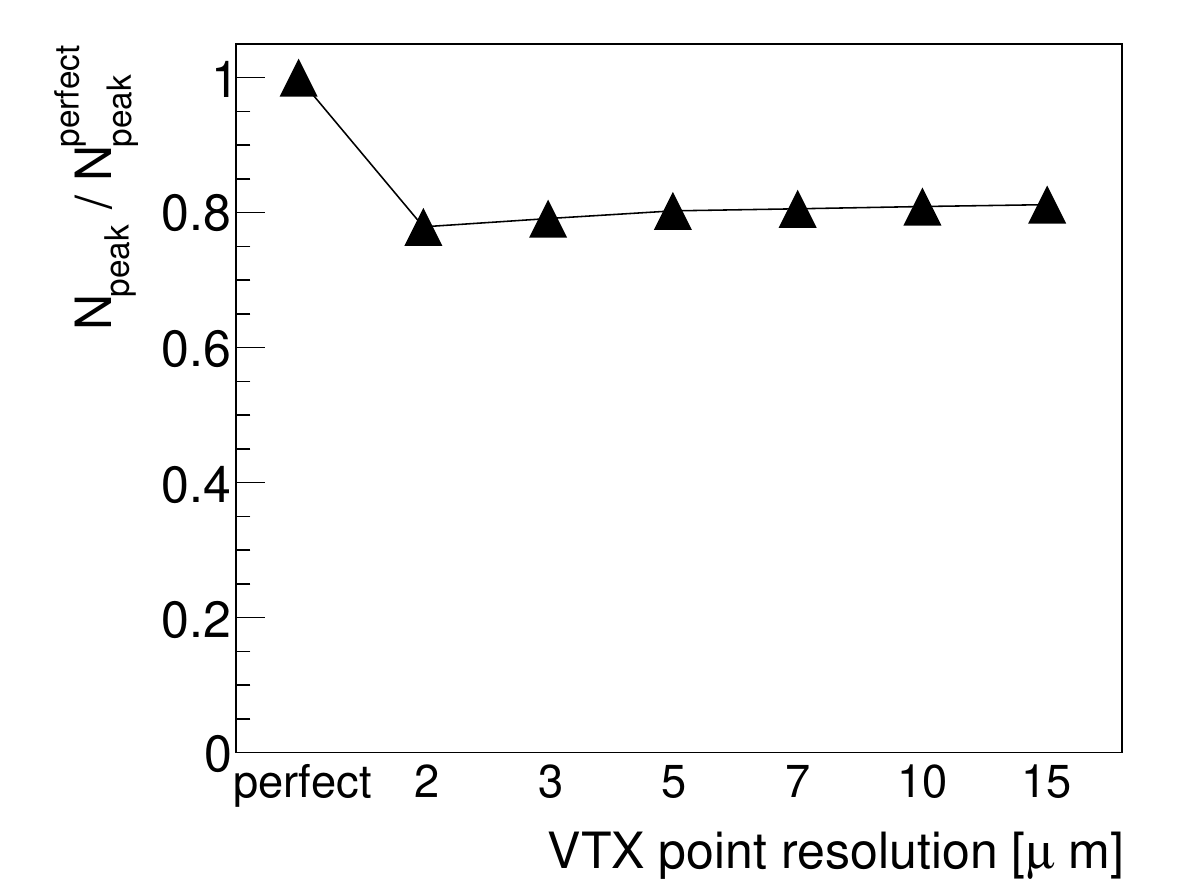} 
    \includegraphics[width=0.35\textwidth]{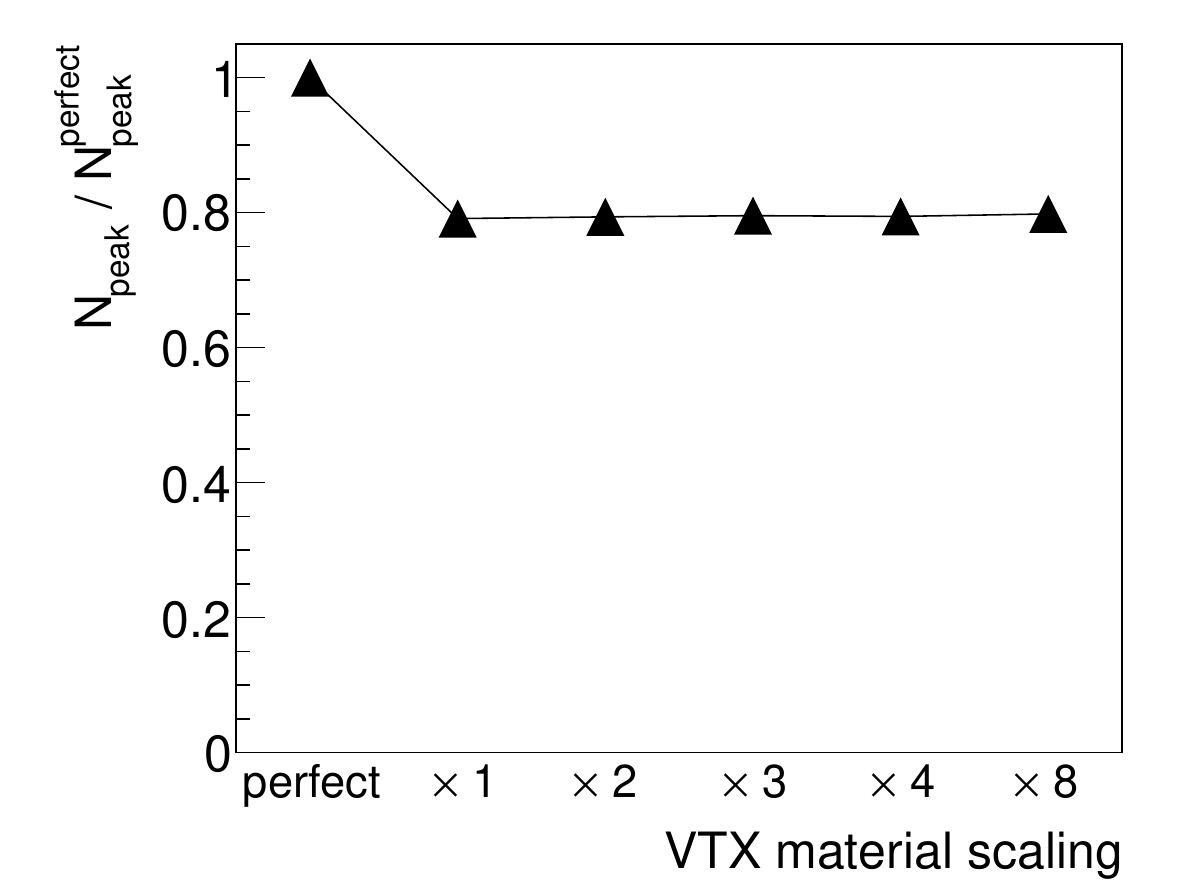} 
    \caption{Dependence of the extraction of the helicity component $k$ on some detector properties,
      for events with $m_{\tau\tau} > 220~\gev$.
      The top two rows show the difference between reconstructed and true values of $k$ (left)
      and the entries in the central peak ($|\delta k| < 0.05$) compared to the case of a perfect detector (right),
      for variations of the photon energy resolution (top) and photon angular resolution (middle row).
      The effect of applying a constrained kinematic fit to photons from $\pi^0$ decay is also shown.
      The lower row shows the central peak entries as a function of the vertex detector point resolution (left)
      and vertex detector material (right).
      When varying a detector parameter, others are fixed to VTX resolution $3\mu m$, VTX material $\times 1$,
      photon energy resolution $15\%/\sqrt{E}\oplus1\%$, and photon angular resolution 0~mrad.
    }
  \end{center}
  \label{fig:compsgv}
\end{figure*}

The resolution on the $k$ component of the polarimeter for these different assumed detector performances are shown in
Fig.~\ref{fig:compsgv}.
The vertex detector material budget and single point resolution have almost no effect,
suggesting that the impact parameter can sufficiently well measured for the purposes
of this analysis even with a conservatively designed vertex detector. 

The photon energy resolution shows only a very limited degradation in the quality of event reconstruction when varying
between an ECAL which perfectly measures photon energies (``0\%'') and one with
an energy resolution typical of the high-granularity sampling calorimeters designed for particle flow reconstruction
($\sigma_E / E = 15\% \oplus 1\%$).

In contrast, the photon angular resolution has a strong effect.
We observe a significant decrease in the efficiency to find a reasonable solution with larger angular smearing.
Increasing the photon smearing from 0 to 1~mrad results in a loss of around $\sfrac{3}{4}$ of events in the central peak.

The narrow $\pi^0$ resonance provides a strong constraint on the kinematics of final states
relying on the ECAL performance
(the $\rho$ and $a_1$ resonances are much wider, and therefore expected to be much less useful).
We have tested the effect of a constrained kinematic fit of the two photons from neutral pion decay
as a method to compensate for finite ECAL resolution in energy and direction.
The photon energies and directions are allowed to vary according to the simulated resolutions, while
imposing the neutral pion invariant mass.
Such a fit has only a small effect on the tau resolution efficiency, as shown in Fig.~\ref{fig:compsgv}.

These observations suggest that for this measurement, an ECAL with excellent position resolution is more critical
than one with excellent energy measurement.
An angular resolution of around 0.1~mrad would avoid too much loss of performance. This corresponds to
a position resolution of $(1\sim2)$~mm on the ECAL face at a typical radius of $\sim 1.7$~m. 
An overly-simplistic estimate suggests that the transverse segmentation of the ECAL should be no larger than
$\sqrt{12} \times (1 \sim 2)~\textrm{mm} = (3.4 \sim 7)~\textrm{mm}$.
A study based on full simulation and reconstruction would be needed to
arrive at more robust requirements on ECAL performance.
Other aspects of a full analysis not considered in the current study,
in particular the correct identification of the tau decay mode, are also
expected to profit from an ECAL with excellent imaging capabilities able to
resolve nearby photons within the tau jet.

\section{Conclusions and outlook}

Quantum tomography of the process $e^+ e^- \to \tau^+ \tau^- (\gamma)$ at high energy colliders
will allow tests of the SM and QM.
The final state is kinematically under-constrained due to the presence of neutrinos in the final state and unseen ISR.
In this paper, we have described a method which provides a weighted set of possible event solutions,
and used these to extract the information required for quantum tomography.
We have estimated the precision with which polarimeter components and their correlations can be extracted,
both for a detector with perfect resolution and for detector models with more realistic assumptions.
The reconstruction of the spin state seems sufficiently precise in the case of a perfect detector. 
Considering more realistic detector performance,
we find that achieving a photon direction resolution of around 0.1~mrad is the most crucial aspect for this analysis,
while the photon energy resolution and vertex detector performance are less impactful.

This analysis was so far performed using fast detector simulation at a CoM energy of 250~\gev.
Other energies differ mostly in the energy and therefore collimation of the tau jets,
and in the amount of ISR. We expect that analysis at the Z pole should be simpler,
given the smaller boost and the reduced effect of ISR.
A more detailed detector simulation and event reconstruction is needed to verify the
conclusions presented here. In particular, the ability to efficiently distinguish
individual photons and pions within a tau jet, to correctly pair photons into neutral pions,
and to identify the tau decay mode are essential aspects of this analysis which
require detailed detector simulations and real reconstruction algorithms to properly quantify.

\section*{Acknowledgements}

I thank my colleagues in the ILD software group for generating the MC samples used in this study,
M.~Berggren for assistance with running SGV,
and A.~Korchin for pointing out an error in the form of the $\taurho$ polarimeter
(Eq.~\ref{eqn:polar}) given in an earlier version of this paper.

\end{document}